\documentclass[twocolumn,prb,showpacs,superscriptaddress]{revtex4}

\usepackage{picinpar}
\usepackage{floatflt}
\usepackage{array}
\usepackage{subfigure}
\usepackage{amsmath}
\usepackage{multirow}
\usepackage[dvips]{graphicx}
\usepackage{amssymb}

\begin{document}

\title{Koopmans' condition for density-functional theory}

\author{Ismaila Dabo}
\email{daboi@cermics.enpc.fr}
\affiliation{Universit\'e Paris-Est, CERMICS, Projet Micmac ENPC-INRIA, 
6-8 avenue Blaise Pascal, 77455 Marne-la-Vall\'ee Cedex 2, France}
\author{Andrea Ferretti}
\affiliation{Department of Materials Science and Engineering,
Massachusetts Institute of Technology, Cambridge, MA, USA}
\altaffiliation[present address: ]{Department of Materials, University of Oxford, Parks Road, Oxford OX1 3PH, United Kingdom}
\author{Nicolas Poilvert}
\affiliation{Department of Materials Science and Engineering,
Massachusetts Institute of Technology, Cambridge, MA, USA}
\author{Yanli Li}
\affiliation{Department of Physics, Institute of Theoretical Physics and Astrophysics, and Fujian Key Lab of Semiconductor
Materials and Applications, Xiamen University, Xiamen 361005, Republic of China}
\author{Nicola Marzari}
\affiliation{Department of Materials Science and Engineering,
Massachusetts Institute of Technology, Cambridge, MA, USA}
\altaffiliation[present address: ]{Department of Materials, University of Oxford, Parks Road, Oxford OX1 3PH, United Kingdom}
\author{Matteo Cococcioni}
\affiliation{Department of Chemical Engineering and Materials Science,
University of Minnesota, Minneapolis, MN, USA}


\begin{abstract}
In approximate Kohn-Sham density-functional theory, self-interaction 
manifests itself as the dependence of the energy of an orbital on 
its fractional occupation. This unphysical behavior
translates into qualitative and quantitative errors that pervade many
fundamental aspects of density-functional predictions.
Here, we first examine self-interaction in
terms of the discrepancy between total and partial electron removal energies,
and then highlight the importance of imposing the generalized Koopmans' condition --- that 
identifies orbital energies as opposite total electron removal energies ---
to resolve this discrepancy.  In the process, we derive
a correction to approximate functionals that, in the frozen-orbital approximation,
eliminates the unphysical occupation dependence of orbital energies up 
to the third order in the single-particle 
densities. This non-Koopmans correction
brings physical meaning to single-particle energies; when applied to common local or semilocal
density functionals it provides results that are
in excellent agreement with experimental data --- with an accuracy comparable to that of GW 
many-body perturbation theory --- while providing an explicit total energy functional 
that preserves or improves on the description of established structural properties. 
\end{abstract}

\maketitle

\section{Introduction}

Density-functional approximations, \cite{PayneTeter1992} which account
for correlated electron interactions via an explicit functional
$E_{\rm xc}$ of the electronic density $\rho$, provide very good predictions of total
energy differences for systems with non-fractional occupations. \cite{ParrYang1989}
One of the notable successes of local and semilocal density-functional calculations is the accurate 
description of ionization processes involving the complete filling or entire depletion of frontier orbitals.
In quantitative terms, the local-spin-density (LSD) approximation and semilocal generalized-gradient approximations (GGAs)
predict the energy differences of such reactions, namely,
the electron affinity
\begin{equation}
A_N=E_N-E_{N+1}
\end{equation}
and the first ionization potential
\begin{equation}
I_N=E_{N-1}-E_N
\label{IPDefinition}
\end{equation}
(where $E_N$ stands for the ground-state energy of the $N$-electron system), with
a precision of a few tenths of an electron-volt, which compares favorably to that of 
more expensive wavefunction methods. \cite{GollErnst2009}

\begin{figure}[ht!]
\includegraphics[width=8cm]{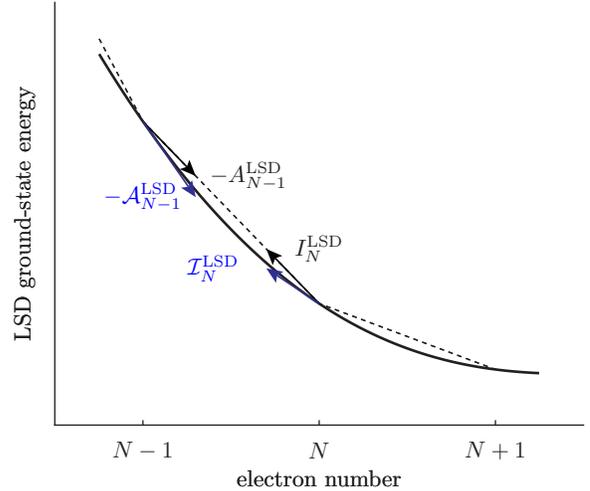}
\caption{Convexity of the LSD ground-state energy as a function of the electron number,
which results from the discrepancy between total and partial differential electron removal energies,
i.e., ${\cal I}^{\rm LSD}_N < I^{\rm LSD}_N=A^{\rm LSD}_{N-1} < {\cal A}^{\rm LSD}_{N-1}$.
\label{PiecewiseLinearity}}
\end{figure}

Considering the excellent performance of density-functional approximations 
in predicting total ionization energies, it is surprising to discover
that the same theories fail in describing partial ionization processes.
As a matter of fact, local and semilocal functionals
overestimate (by as much as 40\%) the absolute energy difference per electron
\begin{equation}
{\cal A}_N=-\left.\frac{dE_M}{dM}\right|_{M=N^+}
\end{equation}
of an infinitesimal electron addition
and underestimate with the same error differential energy changes 
\begin{equation}
{\cal I}_N=-\left.\frac{dE_M}{dM}\right|_{M=N^-}
\end{equation}
upon electron removal.

Analytically, the discrepancy between total and partial differential ionization energies 
manifests itself into the convexity of $E^{\rm LSD}_N$ as a function of $N$ (Fig.~\ref{PiecewiseLinearity})
while the exact ground-state energy $E_N$ versus $N$ is known to be described by
a series of straight-line segments with positive derivative discontinuities 
at integer values of $N$. \cite{PerdewParr1982} (For
simplicity, we restrict the entire discussion to the case of the LSD functional; semilocal GGA functionals exhibit identical trends.)

It is important to note that the discrepancy described above
can be also related to the incorrect analytical behavior of the LSD chemical potential $\mu^{\rm LSD}_N$,
i.e., the Lagrange multiplier associated to particle-number conservation
in the grand-canonical minimization of the total energy,
$\min_\rho \left\{ E[\rho] - \mu \int d{\bf r} \rho({\bf r}) \right\}$, \cite{ParrDonnelly1978,PerdewParr1982,
MarchPucci1983,YangParr1984,PerdewLevy1997} which
 can be interpreted physically as the opposite electronegativity of the system \cite{ParrDonnelly1978,MarchPucci1983}
\begin{equation}
\mu_N= - \chi_N= - \frac{{\cal A}_N+{\cal I}_N}{2},
\end{equation}
and determines the direction and magnitude of electron transfer between separated molecular fragments.\cite{PerdewParr1982}
In fact, the exact dependence of the chemical potential $\mu_N$ when $N$ varies through an integer particle number $Z$ is known to be
\begin{equation}
\mu_N = \left\{ 
\begin{array}{ll}
- I_Z, \; \; \;  Z-1 < N < Z, \\ \\
- \chi_Z^{\cal M} = - \frac 12 (A_Z+I_Z), \; \; \;  N = Z, \\ \\ 
- A_Z, \; \; \;  Z < N < Z + 1, 
\end{array}
\right.
\end{equation}
where $\chi_N^{\cal M}$ denotes the Mulliken electronegativity\cite{Mulliken1934} 
of the $N$-electron system.\cite{PerdewParr1982,PerdewLevy1997}

In related physical terms, the discrepancy between ionization energies is often interpreted 
as arising from electron self-interaction that causes the removal of a small fraction of an electron from a filled
electronic state to be energetically less costly, in absolute value, than its addition to the corresponding empty state. \cite{PerdewZunger1981}

This self-interaction error is at the origin of important quantitative and qualitative failures that 
pervade crucial aspects of
electronic-structure predictions. \cite{CohenMori-Sanchez2008b}
Consider, for example, the dissociation of a cation dimer X$_2^+$ in the infinite interatomic separation limit.
Here, the energy cost of removing a small electron fraction from X is lower than the energy 
gained in adding an electron fraction to X$^+$. As a result, a portion of the electronic charge
will transfer from X to X$^+$, eventually leading to the split-charge configuration 2X$^{\frac 12+}$, 
with a total energy that is correspondingly overstabilized with respect to the exact solution (given, we note in
passing, by any linear combination of the two orthogonal ground states where the electron resides on either
of the two ions).

Related self-interaction artifacts explain other important failures of
local and semilocal functionals in predicting electron-transfer processes, \cite{SitCococcioni2006}
electronic transport, \cite{ToherFilippetti2005} molecular adsorption, 
\cite{KresseGil2003,DaboWieckowski2007} reaction barriers and energies, \cite{KulikCococcioni2006,Mori-SanchezCohen2006a,
VydrovScuseria2007,JohnsonMori-Sanchez2008} and electrical polarization in extended systems.
\cite{KummelKronik2004,BaerNeuhauser2005,UmariWillamson2005,RuzsinszkyPerdew2008}
Self-interaction is also connected to the underestimation of the differential energy gap
$\epsilon^{\rm deriv}_{\rm gap}={\cal I}_N-{\cal A}_N$
of molecules, semiconductors, and insulators within LSD and GGAs. 
\cite{PerdewLevy1983,CohenMori-Sanchez2008a,Mori-SanchezCohen2008,KummelKronik2008,CohenMori-Sanchez2009}

This work is organized as follows. First, we reexamine the Perdew-Zunger one-electron self-interaction correction in terms
of total and differential ionization energies. Then, after deriving an exact measure of the unphysical curvature
of $E_N$ as a function of $N$ based
on the generalized Koopmans' theorem, we introduce a functional that minimizes self-interaction errors by enforcing Koopmans' condition, 
thereby largely eliminating the discrepancy between total and differential ionization 
energies while restoring the piecewise linearity of the ground-state energy $E_N$ as a function of 
the number of electrons $N$. We conclude the study with extensive atomic and molecular calculations
to demonstrate the predictive power of the non-Koopmans self-interaction correction.

\section{Self-interaction correction}

\subsection{Perdew-Zunger one-electron correction}

\label{PZSection}

Several methods have been proposed to eliminate self-interaction contributions and restore the internal consistency between 
total and partial electron removal energies. \cite{HeatonPederson1987,FilippettiSpaldin2003,CococcionideGironcoli2005,DAvezacCalandra2005,
RuzsinszkyPerdew2006,AnisimovKozhevnikov2007,FerreiraMarques2008,KummelKronik2008,StengelSpaldin2008,LanyZunger2009,CampoCococcioni2010,LanyZunger2010}
A widely used approach is that introduced by Perdew and Zunger, \cite{PerdewZunger1981} which consists of correcting self-interaction
in the one-electron approximation by subtracting one-electron Hartree and
exchange-correlation contributions. Explicitly, at the LSD level, the Perdew-Zunger (PZ) orbital-dependent functional
and Hamiltonian are defined as 
\begin{eqnarray}
E^{\rm PZ} & = & E^{\rm LSD} + \sum_{i\sigma} \left( - E_{\rm H}[\rho_{i\sigma}] - E_{\rm xc}[\rho_{i\sigma}] \right) \\
           & = & E^{\rm LSD} + \sum_{i\sigma} \Xi_{i\sigma}
\end{eqnarray}
\begin{equation}
\hat h^{\rm PZ}_{i\sigma} = \hat h^{\rm LSD}[\rho] - \hat v_{\rm H}[\rho_{i\sigma}]-\hat v_{\rm xc,\sigma}[\rho_{i\sigma}],
\end{equation}
where $\rho_{i\sigma}({\bf r})$ denotes the orbital density ($\sigma=\pm\frac{1}{2}$ represents the spin), 
$v_{\rm H}({\bf r})=\delta E_{\rm H} / \delta \rho({\bf r})$ 
is the electrostatic Hartree potential, $v_{\rm xc,\sigma}({\bf r})=\delta E_{\rm xc} / \delta \rho_\sigma({\bf r})$ 
stands for the spin-dependent exchange-correlation potential, and the PZ one-electron corrective energy contributions
are defined as \linebreak $\Xi_{i\sigma}=- E_{\rm H}[\rho_{i\sigma}] - E_{\rm xc}[\rho_{i\sigma}]$.

\begin{figure}[ht!]
\includegraphics[width=8cm]{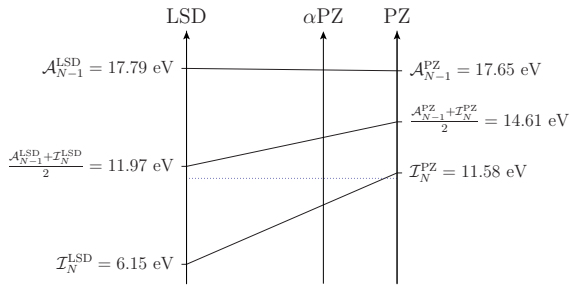}
\caption{LSD differential ionization energies ${\cal A}^{\rm LSD}_{N-1}$,
${\cal I}^{\rm LSD}_N$, and their arithmetic mean $\frac{1}{2}({\cal A}^{\rm LSD}_{N-1}+{\cal I}^{\rm LSD}_N)$ 
compared with PZ differential ionization energies
for carbon. The dotted line indicates the position of the experimental ionization potential. The behavior
of a PZ functional linearly downscaled by a factor $\alpha^{\rm PZ}$ is also shown.
\label{PZAdmixture}}
\end{figure}

The effect of the PZ self-interaction correction on the partial electron removal
energies of an isolated carbon atom is depicted in Fig.~\ref{PZAdmixture}. 
We observe that the LSD differential ionization potential ${\cal I}_N^{\rm LSD}$
and electron affinity ${\cal A}_{N-1}^{\rm LSD}$ deviate by more than 5 eV
from the experimental total removal energy, whereas their average 
is in close agreement with experiment. The PZ correction improves 
the precision of the predicted ionization potential ${\cal I}^{\rm PZ}_N$, reducing the error 
to less than 0.4 eV. The inaccuracy of
the PZ electron affinity ${\cal A}^{\rm PZ}_{N-1}$
remains unchanged due to the fact that the PZ one-electron correction vanishes for empty states 
(the slight variation of the differential electron affinity from 17.79 to 17.65 eV
is only due to the self-consistent reconfiguration of the occupied manifold).
Consequently, the PZ functional achieves a substantial but only partial correction of self-interaction,
reducing the unphysical convexity, i.e., the discrepancy ${\cal A}_{N-1}-{\cal I}_N$, by a factor of $\frac 12$. 

Another notable feature in Fig.~\ref{PZAdmixture} is the overestimation of the 
average of the ionization potential ${\cal I}^{\rm PZ}_N$ and electron affinity ${\cal A}^{\rm PZ}_{N-1}$.
Since the average $\frac{1}{2}({\cal A}^{\rm PZ}_{N-1}+{\cal I}^{\rm PZ}_N)$
approximates $I^{\rm PZ}_N$, \cite{PerdewZunger1981,Bruneval2009}
this deviation translates into overestimated total ionization energies
and excessively negative ground-state energies $E^{\rm PZ}_N=-\sum_{M=1}^N I^{\rm PZ}_M$. 
(Semi-empirical relations between total energies and low-order ionization potentials 
were also evidenced by Pucci and March in Ref.~[\onlinecite{PucciMarch1982}].)
In the case of isolated atoms, the energy underestimation improves (fortuitously) on the description of electron correlation,
bringing PZ ground-state energies in remarkable agreement with experiment. \cite{PerdewZunger1981}
However, in the more general case of many-electron polyatomic systems,
PZ ground-state energies are found to be overcorrelated, yielding
inaccurate dissociation energies and excessively short bond lengths. \cite{GoedeckerUmrigar1997}

Various downscaling methods have been proposed to correct the above trends. Nevertheless,
the performance of such schemes is inherently limited by the fact that downscaling the PZ correction
impairs the accuracy of differential ionization energies. 
For example, it is seen in Fig.~\ref{PZAdmixture}
that a functional $E^{\alpha \rm PZ}$, in which the self-interaction correction is linearly downscaled
by a factor $\alpha^{\rm PZ}$,
\begin{equation}
E^{\alpha \rm PZ}=E^{\rm LSD}+\alpha^{\rm PZ} (E^{\rm PZ}-E^{\rm LSD})
\end{equation}
cannot counterbalance the deviation of \linebreak $I^{\rm PZ}_N \approx \frac{1}{2}({\cal A}^{\rm PZ}_{N-1}+{\cal I}^{\rm PZ}_N)$
without altering the precision of ${\cal I}^{\rm PZ}_N$. 
More sophisticated downscaling methods exhibit identical trends. \cite{RuzsinszkyPerdew2006,
VydrovScuseria2006a,VydrovScuseria2006b,RuzsinszkyPerdew2007}

In the next sections, we will derive a correction
that cancels self-interaction errors for systems with fractional occupations
without affecting the precision of total energy differences and equilibrium structural properties 
for systems with non-fractional occupations.

\subsection{Measure of self-interaction}

The initial conceptual step in the construction of the self-interaction correction is
to set forth a quantitative measure of self-interaction errors
valid in the many-electron case --- i.e., 
beyond the one-electron approximation of the Perdew-Zunger correction.
With this self-interaction measure in hand, we will construct
an improved self-interaction correction functional,
working first in the simplified picture where electronic orbitals are kept frozen (Sec. \ref{BareNKSection}),
and then considering orbital relaxation as the next stage of refinement (Sec. \ref{ScreenedNKSection}).

Before doing so, to derive this measure, we start from the physical intuition that self-interaction 
relates to the unphysical variation of the energy of an orbital $\epsilon_{i\sigma}(f)$ as a function of its own occupation $f$.
Hence, a {\it necessary} non-self-interaction condition can be written as 
\begin{equation}
\left. \frac{d \epsilon_{i\sigma}(f')}{d f'} \right|_{f'=f} =0, \; \; \;  0 \le f \le 1.
\label{EpsilonCondition}
\end{equation}
(Here, $f$ stands for a variable occupation while $f_{i\sigma}$ 
denotes the orbital occupation that enters into the expression of the total energy functional.)
Invoking Janak's theorem,\cite{Janak1978} this necessary condition on orbital-energy derivatives can be restated as
a criterion on the curvature of the total energy:
\begin{equation}
\left. \frac{d^2 E_{i\sigma}(f')}{d f'^2} \right|_{f'=f}=0, \; \; \;  0 \le f \le 1,
\label{SecondDerivativeCondition}
\end{equation}
where $E_{i\sigma}(f)$ is the total energy $E$ minimized under the constraint $f_{i\sigma}=f$ (leaving all other occupations unchanged).
Thus, in the absence of self-interaction, the total energy 
does not display any curvature upon varying occupations. 
As a corollary, any self-interaction-free functional
satisfies the linearity condition
\begin{equation}
\Delta E_{i\sigma} = - f_{i\sigma} \left. \frac{d E_{i\sigma}(f')}{d f'}\right|_{f'=f}, \; \; \;  0 \le f \le 1,
\end{equation}
where 
\begin{equation}
\Delta E_{i\sigma} = E_{i\sigma}(0)-E_{i\sigma}(f_{i\sigma})
\end{equation} 
denotes the removal energy of the orbital $\psi_{i\sigma}$.\cite{RuzsinszkyPerdew2007,CohenMori-Sanchez2008b,Mori-SanchezCohen2006b}

Invoking once more Janak's theorem, the above non-self-interaction condition 
yields the {\it generalized Koopmans' theorem},
\begin{equation}
\Delta E_{i\sigma} = - f_{i\sigma} \epsilon_{i\sigma}(f), \; \; \;  0 \le f \le 1,
\label{GeneralizedKoopmansTheorem}
\end{equation}
(The equivalence between Eq.~(\ref{EpsilonCondition}) and Eq.~(\ref{GeneralizedKoopmansTheorem}) 
highlights the significance of Koopmans' theorem for the quantitative assessment of self-interaction.)
It is then convenient to measure self-interaction in terms of 
the energies $\Pi_{i\sigma}$ 
first introduced by Perdew and Zunger (see Sec. IID in Ref.~[\onlinecite{PerdewZunger1981}]) in analyzing discrepancies between orbital energies
and vertical ionization energies: 
\footnote{Our definition of the self-interaction energy
$\Pi_{i\sigma}(f)$ differs slightly from that of Ref. [\onlinecite{PerdewZunger1981}] in that it is
occupation-dependent and involves relaxed energies, thereby providing
an exact measure of the curvature of the ground-state energy on the entire occupation segment $0 \le f \le 1$.}
\begin{equation}
\Pi_{i\sigma}(f) = f_{i\sigma} \epsilon_{i\sigma}(f) + \Delta E_{i\sigma}.
\label{PiOriginalDefinition}
\end{equation}
Because the self-interaction energies $\Pi_{i\sigma}$ quantify deviations
from the Koopmans linearity [Eq.~(\ref{GeneralizedKoopmansTheorem})], they will be termed here
{\it non-Koopmans energies} --- the same terminology as that employed in Refs.~[\onlinecite{PerdewZunger1981}]
and [\onlinecite{HeatonPederson1987}].

Making then use of Slater's theorem,\cite{Slater1974}
\begin{equation}
\Delta E_{i\sigma}=- \int^{f_{i\sigma}}_0 \epsilon_{i\sigma} (f) df,
\end{equation}
the non-Koopmans energy can be rewritten as  
\begin{equation}
\Pi_{i\sigma}(f) = \int_0^{f_{i\sigma}} \Big[\epsilon_{i\sigma} (f) - \epsilon_{i\sigma}(f')\Big]df'. 
\label{PiIntegralDefinition}
\end{equation}
In this form, the energy $\Pi_{i\sigma}(f)$ is clearly seen to correspond to the integrated change of the orbital energy 
upon varying the orbital occupation --- in particular, 
it is straightforward to verify that $\Pi_{i\sigma}(f)=0$
if the orbital energy $\epsilon_{i\sigma}(f')$ does not vary with the orbital occupation $f'$.

Using the measure defined in Eq.~(\ref{PiOriginalDefinition}), 
the non-self-interaction criterion [Eq.~(\ref{EpsilonCondition})] can be restated exactly as a {\it generalized Koopmans' condition},
\begin{equation}
\Pi_{i\sigma}(f)=0, \; \; \;  0 \le f \le 1.
\label{KoopmansCondition}
\end{equation}
Equation (\ref{KoopmansCondition}) is of central importance in this work;
it provides a simple quantitative criterion in terms of a rigorous energy-nonlinearity measure
to assess and correct self-interaction errors.
(The suggestion of using non-Koopmans corrections to minimize self-interaction
has been recently introduced, in preliminary form, by Dabo, Cococcioni, and Marzari
 in Ref.~[\onlinecite{DaboCococcioni2009}],
and, in heuristic form, by Lany and Zunger in Ref.~[\onlinecite{LanyZunger2009}].)
The consequences of Koopmans' condition are discussed in the next sections.

\subsection{Bare non-Koopmans correction}

\label{BareNKSection}

On the basis of the above quantitative analysis, our objective now is to 
linearize the dependence of the total energy as a function of orbital occupations by modifying
the expression of the energy functional in order to cancel the non-Koopmans terms $\Pi_{i\sigma}(f)$.

To render this complex problem tractable, we first consider the restricted case where all orbitals are frozen while the occupation
of one of them is changing in the course of a fictitious ionization process (the frozen-orbital approximation). Within
this paradigm, Eq.~(\ref{KoopmansCondition}) becomes the {\it restricted Koopmans' condition},
\begin{equation}
\Pi_{i\sigma}^u(f)=0, \; \; \;  0 \le f \le 1,
\label{RestrictedKoopmansCondition}
\end{equation}
where the superscript $u$ stands for {\it unrelaxed}.
Here, the frozen-orbital non-Koopmans energy $\Pi_{i\sigma}^u$ is defined as
\begin{equation}
\Pi_{i\sigma}^u(f) = \int_0^{f_{i\sigma}} df' 
\Big[ \epsilon_{i\sigma}^u (f) - \epsilon_{i\sigma}^u (f') \Big],
\label{PiEnergy}
\end{equation}
where $\epsilon_{i\sigma}^u(f)$
is the unrelaxed orbital energy calculated keeping all the orbitals frozen
while setting $f_{i\sigma}$ to be $f$. 
We underscore that in the specific case of one-electron systems and for $f=0$,
Eq.~(\ref{RestrictedKoopmansCondition}) yields the one-electron self-interaction
condition of Perdew and Zunger,\cite{PerdewZunger1981}
\begin{equation}
\Xi_{i\sigma} = 0.
\end{equation}
The restricted Koopmans' condition is thus seen to encompass the Perdew-Zunger condition,
thereby representing a more comprehensive criterion for assessing and correcting self-interaction errors.

Expectedly, satisfying the restricted Koopmans' condition is exactly equivalent
to fulfilling the {\it restricted Koopmans' theorem},
\begin{equation}
\Delta E_{i\sigma}^u = - f_{i\sigma} \epsilon_{i\sigma}^u(f),
\; \; \;  0 \le f \le 1,
\label{KoopmansTheorem}
\end{equation}
where
\begin{equation}
\Delta E_{i\sigma}^u=E_{i\sigma}^u(0)-E_{i\sigma}^u(f_{i\sigma})
\end{equation} 
denotes the unrelaxed electron removal energy.

An example of a functional that exhibits a linear frozen-orbital energy dependence
is provided by the Hartree-Fock (HF) theory, generalized to fractional occupations. Indeed, at the HF level,
one can verify that
\begin{equation}
\Pi_{i\sigma}^{u, \rm HF}(f)=0, \; \; \;  0 \le f \le 1,
\label{HFKoopmansCondition}
\end{equation}
due to the fact that the expectation value 
$\epsilon^{u, \rm HF}_{i\sigma}(f)$ does not depend on $f$.

For functionals that do not satisfy the restricted Koopmans' condition [Eq.~(\ref{RestrictedKoopmansCondition})], 
the unrelaxed electron removal energy can only be expressed in terms of the restricted Slater integral,
\begin{equation}
\Delta E^u_{i\sigma} =- \int^{f_{i\sigma}}_0 df \epsilon^u_{i\sigma}(f),
\label{SlaterEnergy}
\end{equation}
(note that this relation is satisfied by any functional whether it is subject or not to self-interaction errors).

These considerations allow us to introduce our corrected functional
aiming to satisfy Koopmans' condition; at frozen orbital, it is obtained by
replacing Slater terms --- Eq.~(\ref{SlaterEnergy}), where single-particle energies are function
of their own occupation --- with Koopmans terms --- Eq.~(\ref{KoopmansTheorem}), where single-particle energies
do not depend on occupations. In particular, we evaluate here the Koopmans terms at a given orbital occupation
$f=f_{\rm ref}$ that defines the reference transition state (the
determination of the reference occupation $f_{\rm ref}$ shall be explained
on the basis of Slater's approximation and exchange-correlation hole arguments in Sec. \ref{ReferenceOccupationSection}). 
Explicitly, in the case of the LSD functional,
the non-Koopmans (NK) self-interaction correction to the energy is defined as
\begin{equation}
E^{\rm NK}=E^{\rm LSD} - \sum_{i\sigma} \Big[ - f_{i\sigma} \epsilon_{i\sigma}^u(f_{\rm ref}) 
+\int^{f_{i\sigma}}_0 df \epsilon^u_{i\sigma} (f) \Big],
\label{NonKoopmansFunctional}
\end{equation}
where the negative sign in front of the sum follows from the convention that ionization energies are positive.
Rewriting Eq.~(\ref{NonKoopmansFunctional}) in terms of the frozen-orbital non-Koopmans energies of
Eq.~(\ref{PiEnergy}), we obtain
\begin{equation}
E^{\rm NK}=E^{\rm LSD} + \sum_{i\sigma} \Pi^{u,\rm LSD}_{i\sigma}(f_{\rm ref}),
\label{NonKoopmansFunctional2}
\end{equation}
where the non-Koopmans corrective terms can now be recast into the explicit functional form
\begin{widetext}
\begin{eqnarray}
\label{PiDefinition}
\Pi_{i\sigma}^{u, \rm LSD}(f_{\rm ref}) = f_{i\sigma}(2 f_{\rm ref}-f_{i\sigma} ) E_{\rm H}\left[n_{i\sigma}\right] 
- E_{\text{xc}}^{\rm LSD}[\rho]
+  E_{\text{xc}}^{\rm LSD}[\rho-\rho_{i\sigma}] + \int d\mathbf{r} \rho_{i\sigma}(\mathbf{r})
v^{\rm LSD}_{\text{xc},\sigma}(\mathbf{r};[\rho_{i\sigma}^{\text{ref}}]) d\mathbf{r},
\end{eqnarray}
\end{widetext}
where $n_{i\sigma}(\mathbf{r})=|\psi_{i\sigma}|^2(\mathbf{r})$ and $\rho_{i\sigma}(\mathbf{r})=f_{i\sigma}n_{i\sigma}(\mathbf{r})$.
Here, the electronic density $\rho_{i\sigma}^{\rm ref}(\mathbf{r})$ stands for the reference transition-state density 
\begin{eqnarray}
 \rho_{i\sigma}^{\rm ref}(\mathbf{r}) &=& f_{\rm ref} n_{i\sigma}(\mathbf{r})
    + \sum_{j\sigma' \neq i\sigma} f_{j\sigma'} n_{j\sigma'}(\mathbf{r})
\label{ReferenceDensity}
\\
    &=& \rho(\mathbf{r}) + (f_{\rm ref} -f_{i\sigma})\, n_{i\sigma}(\mathbf{r}).
\end{eqnarray}

We now derive the expression of the orbital-dependent NK Hamiltonian. To this end, we first calculate the functional derivative
of the energy contribution $\Pi^{u, \rm LSD}_{i\sigma}(f_{\rm ref})$ with respect 
to variations of the orbital density $\rho_{i\sigma}$.
The expression of the functional derivatives reads 
\begin{multline}
\frac{\delta \Pi_{i\sigma}^{u,\rm LSD}}{\delta \rho_{i\sigma}({\bf r})} 
 = (f_{\rm ref} - f_{i\sigma}) v_{\rm H}({\bf r};\left[n_{i\sigma}\right]) \\
+ v^{\rm LSD}_{\rm xc,\sigma}({\bf r};[\rho_{i\sigma}^{\rm ref}])
-  v^{\rm LSD}_{\rm xc, \sigma}({\bf r};[\rho]) + w^{\rm LSD}_{{\rm ref},i \sigma}({\bf r}).
\label{StraightDerivative}
\end{multline}
In Eq.~(\ref{StraightDerivative}), the potential $w^{\rm LSD}_{{\rm ref},i \sigma}$ denotes
\begin{eqnarray}
w^{\rm LSD}_{{\rm ref},i \sigma}({\bf r}) = f_{\rm ref} \Bigg[ \int 
d{\bf r}' f^{\rm LSD}_{\rm Hxc,\sigma \sigma}({\bf r},{\bf r}';[\rho_{i\sigma}^{\rm ref}]) 
n_{i\sigma}({\bf r}') \nonumber \\
- \int d{\bf r}'d{\bf r}'' f^{\rm LSD}_{\rm Hxc, \sigma \sigma}({\bf r}',{\bf r}'';[\rho_{i\sigma}^{\rm ref}]) 
n_{i\sigma}({\bf r}')n_{i\sigma}({\bf r}'') \Bigg]
\label{Wref}
\end{eqnarray}
where $f^{\rm LSD}_{\rm Hxc, \sigma \sigma'}(\mathbf{r},\mathbf{r}')=\delta^2 (E_{\rm H}+E^{\rm LSD}_{\rm xc})/
\delta \rho_\sigma(\mathbf{r})  \delta \rho_{\sigma'}(\mathbf{r}')$
is the second-order functional derivative of the LSD energy. \footnote{In deriving Eqs.~(\ref{StraightDerivative}) and (\ref{Wref}), 
orbital occupations and squared wavefunctions must be written explicitly in terms of orbital densities
as $f_{i\sigma}=\int d{\bf r}' \rho_{i\sigma}({\bf r}') $ and 
$n_{i\sigma}({\bf r})=\rho_{i\sigma}({\bf r})/\int d{\bf r}' \rho_{i\sigma}({\bf r}')$.}
Focusing then on the cross derivatives, we obtain
\begin{multline}
\frac{\delta \Pi_{j\sigma'}^{u,\rm LSD}}{\delta \rho_{i\sigma}({\bf r})} 
= v^{\rm LSD}_{\rm xc,\sigma}({\bf r};[\rho - \rho_{j\sigma'}])
- v^{\rm LSD}_{\rm xc,\sigma}({\bf r};[\rho]) \\
+ \int d{\bf r}' f^{\rm LSD}_{\rm xc,\sigma \sigma'}({\bf r},{\bf r}';[\rho_{j\sigma'}^{\rm ref}]) 
\rho_{j \sigma'}({\bf r}'),
\end{multline}
where $f^{\rm LSD}_{\rm xc, \sigma \sigma'}(\mathbf{r},\mathbf{r}')$ is the exchange-correlation contribution to
$f^{\rm LSD}_{\rm Hxc, \sigma \sigma'}(\mathbf{r},\mathbf{r}')$.
As a final result, the orbital-dependent NK Hamiltonian can be cast into the form
\begin{equation}
\hat h_{i\sigma}^{\rm NK} =
\hat h^{\rm LSD}[\rho_{i\sigma}^{\rm ref}]
+ \hat w^{\rm LSD}_{{\rm ref},i \sigma} + \hat w^{\rm LSD}_{{\rm xd},i \sigma},
\label{NonKoopmansHamiltonian}
\end{equation}
where $w^{\rm LSD}_{{\rm xd},i \sigma}$ stands for the cross-derivative potential
\begin{equation}
w^{\rm LSD}_{{\rm xd}, i \sigma}({\bf r}) = 
\sum_{j\sigma' \neq i \sigma} \frac{\delta \Pi_{j\sigma'}^{u,\rm LSD}}{\delta \rho_{i\sigma}({\bf r})}.
\end{equation}

In a nutshell, the NK Hamiltonian consists of the uncorrected LSD Hamiltonian calculated
at the reference density $\hat h^{\rm LSD}[\rho_{i\sigma}^{\rm ref}]$
with the addition of two variational potentials. The first additional term
$\hat w^{\rm LSD}_{{\rm ref},i \sigma}$ results from the variation of the reference density
as a function of $\rho_{i\sigma}$
while the second term $\hat w^{\rm LSD}_{{\rm xd},i \sigma}$ springs from the cross-dependence
of the non-Koopmans corrective terms. 
The effect of the $\hat w^{\rm LSD}_{{\rm ref},i \sigma}$ and $\hat w^{\rm LSD}_{{\rm xd},i \sigma}$ 
contributions that arise as by-products of variationality is analyzed in the next section.

\subsection{Comparative assessment}

\label{ComparativeAssessmentSection}

\begin{figure}[ht!]
\includegraphics[width=8.5cm]{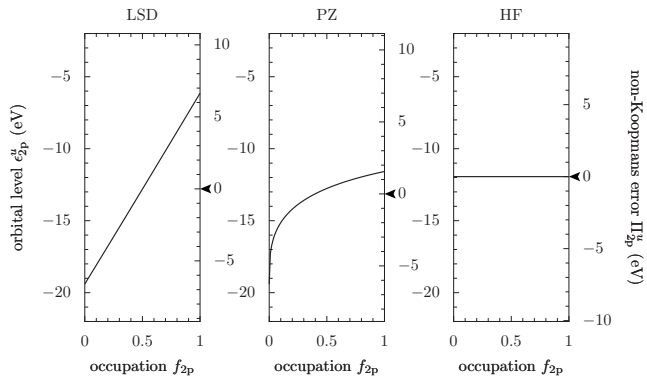}
\caption{LSD, PZ, and HF unrelaxed orbital energies and residual unrelaxed non-Koopmans errors of the highest occupied state of carbon.
The black arrow highlights the zero value for the non-Koopmans error scale.
\label{CarbonEigenvalues}}
\end{figure}

\begin{figure}[ht!]
\includegraphics[width=8.5cm]{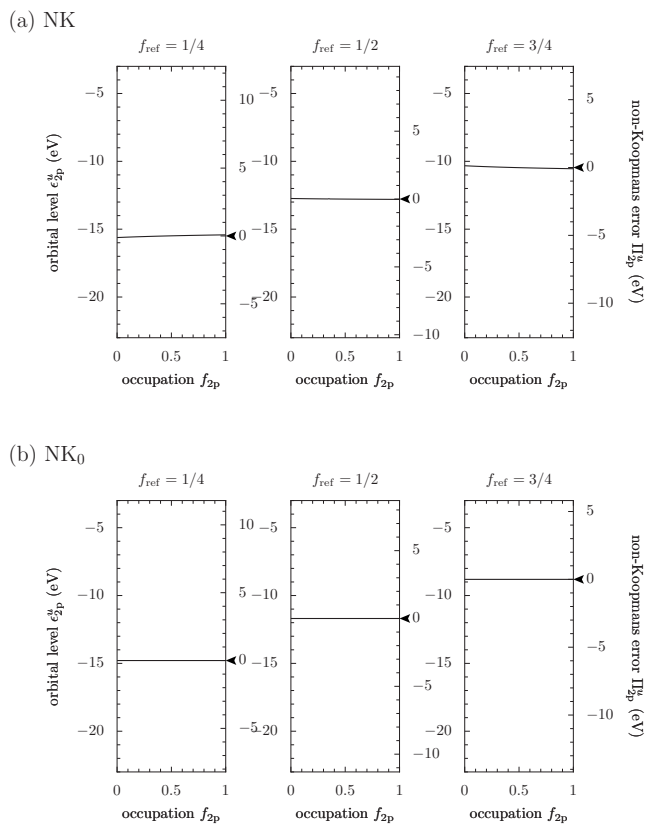}
\caption{Non-Koopmans unrelaxed orbital energies and residual unrelaxed non-Koopmans errors of the highest occupied state of carbon 
for select arbitrary values of the reference occupation ($f_{\rm ref}=\frac{1}{4}$, $\frac{1}{2}$, and $\frac{3}{4}$)
using (a) the NK total-energy method and (b) the NK$_0$ orbital-energy method.
The black arrow highlights the zero value for the non-Koopmans error scale.
\label{CarbonEigenvaluesNKNK0}}
\end{figure}

In this section, we assess the performance of the NK self-interaction correction,
particularly focusing on the effect of variational terms 
on the accuracy of NK orbital predictions --- i.e., on the cancellation of
the unrelaxed frozen-orbital self-interaction measure $\Pi^{u, \rm NK}_{i\sigma}(f)$.

One simple and probably the most direct way to evaluate the influence of 
$\hat w^{\rm LSD}_{{\rm ref},i \sigma}$ and $\hat w^{\rm LSD}_{{\rm xd},i \sigma}$ is to introduce a non-variational 
orbital-energy scheme, the NK$_0$ method, that 
consists of freezing the dependence of the reference transition-state densities
and the cross-dependence of corrective energy terms,
thereby eliminating $\hat w^{\rm LSD}_{{\rm ref},i \sigma}$ 
and $\hat w^{\rm LSD}_{{\rm xd},i \sigma}$ contributions to the effective potential.
Computed NK and NK$_0$ orbital levels can then be compared for the direct assessment of
$\hat w^{\rm LSD}_{{\rm ref},i \sigma}$ and $\hat w^{\rm LSD}_{{\rm xd},i \sigma}$ errors.
Explicitly, the NK$_0$ Hamiltonian can be written as
\begin{equation}
\hat h_{i\sigma}^{\rm NK_0}=\hat h^{\rm LSD}[\rho]+
\left. \frac{\delta \Pi_{i\sigma}^{u, \rm LSD}}{\delta \hat \rho_{i\sigma}} \right|_{\rho_{i\sigma}^{\rm ref}={\rm cst}},
\end{equation}
\begin{equation}
\hat h_{i\sigma}^{\rm NK_0}=\hat h^{\rm LSD}[\rho_{i\sigma}^{\rm ref}].
\label{NonKoopmansZeroHamiltonian}
\end{equation}
In the NK$_0$ optimization scheme,
the Hamiltonian given by Eq.~(\ref{NonKoopmansZeroHamiltonian})
is employed to propagate orbital degrees of freedom at fixed $\rho_{i\sigma}^{\text{ref}}$. 
Reference transition-state densities are then updated according to Eq.~($\ref{ReferenceDensity}$). 
The procedure is iterated until self-consistency.

Due to the loss of variationality, 
the obvious practical limitation of the non-variational NK$_0$ orbital-energy method is that it cannot provide total energies
and interatomic forces.
However, NK$_0$ is of great utility in evaluating the intrinsic performance of the NK correction.
In itself, the NK$_0$ formulation is also useful 
in determining orbital energy properties that are particularly affected by 
$\hat w^{\rm LSD}_{{\rm ref},i \sigma}$ and $\hat w^{\rm LSD}_{{\rm xd},i \sigma}$ errors.

Focusing now on computational predictions,
the occupation dependencies of the LSD, HF, PZ, and NK unrelaxed orbital energies
\begin{equation}
\epsilon^u_{i\sigma}(f)=\left. \frac{d E^u_{i\sigma}(f')}{d f'}\right|_{f'=f}
\end{equation}
of the highest atomic orbital of carbon are depicted in Figs.~\ref{CarbonEigenvalues} and \ref{CarbonEigenvaluesNKNK0}(a). 
The salient feature of the LSD graph is the large variation of the orbital energy from 
$-19.40$ to $-6.15$ eV, reflecting the strong nonlinearity of the corresponding unrelaxed ionization curve.
The PZ variation is found to be twice lower than for LSD, confirming the trends observed
in Sec. \ref{PZSection}. In contrast, the HF unrelaxed ionization curve exhibits a perfectly linear behavior (i.e., the unrelaxed orbital
energy remains constant). This trend is closely reproduced by the NK functional [Fig.~\ref{CarbonEigenvaluesNKNK0}(a)]; 
on the scale of LSD residual non-Koopmans errors, 
the eye can barely distinguish any deviation of the NK unrelaxed orbital energies as a function of $f_{i\sigma}$
regardless of the value of the reference occupation.

The above observation is due to the fact that the variational contribution $\hat w^{\rm LSD}_{{\rm ref},i \sigma}$
affects the orbital energy $\epsilon_{i\sigma}^{\rm NK}$ indirectly, i.e., only through the self-consistent response
of the orbital densities since 
$\langle\psi_{i\sigma}|\hat w^{\rm LSD}_{{\rm ref},i \sigma}|\psi_{i\sigma}\rangle=0$
at self-consistency.
Furthermore, a Taylor series expansion of $\hat w^{\rm LSD}_{{\rm xd},i\sigma}$ reveals that
$\langle \psi_{i\sigma} | \hat w^{\rm LSD}_{{\rm xd},i \sigma} | \psi_{i\sigma} \rangle$ 
does not cause notable departure from the linear Koopmans behavior. In quantitative terms, the dominant 
term in the expansion of the residual NK self-interaction error is of the fourth order in
orbital densities:
\begin{widetext}
\begin{eqnarray}
\Pi^{u, \rm NK}_{i\sigma}(f) 
& = & \frac 14 \sum_{j\sigma'\neq i\sigma} f_{j\sigma'}(2f_{\rm ref}-f_{j\sigma'})(2f-f_{i\sigma})f_{i\sigma} \times \nonumber \\
& \times & \int d {\bf r}_{1234} 
f^{(4),{\rm LSD}}_{{\rm xc},\sigma \sigma' \sigma' \sigma}({\bf r}_{1234};[\rho_{j\sigma'}^{\rm ref}])
n_{i\sigma}({\bf r}_1)n_{j\sigma'}({\bf r}_2)n_{j\sigma'}({\bf r}_3)n_{i\sigma}({\bf r}_4)+ \cdots
\end{eqnarray}
\end{widetext}
(where $f_{{\rm xc},\sigma_{12\cdots n}}^{(n),\rm LSD}({\bf r}_{12\cdots n})$ 
denotes the $n$th order functional derivative of the LSD exchange-correlation energy), 
whereas the PZ correction is found to be less accurate in minimizing the self-interaction measure 
by one order of precision:
\begin{widetext}
\begin{equation}
\Pi^{u, \rm PZ}_{i\sigma}(f) 
= \frac 12 \sum_{j\neq i} (2f-f_{i\sigma})f_{i\sigma}f_{j\sigma}
\int d {\bf r}_{123} f^{(3),{\rm LSD}}_{{\rm xc},\sigma \sigma \sigma}({\bf r}_{123};[\rho-\rho_{i\sigma}])
n_{i\sigma}({\bf r}_1)n_{i\sigma}({\bf r}_2)n_{j\sigma}({\bf r}_3)+ \cdots.
\end{equation}
\end{widetext}

Despite the very good accuracy of the NK correction, direct confrontation with NK$_0$ results 
--- for which the non-Koopmans measure $\Pi_{i\sigma}^{u,\rm NK_0}(f)$ [Eq.~(\ref{PiEnergy})]
is obviously canceled for any value of $f$ --- reveals that NK tends to underestimate
orbital energies with deviations of 0.5 to 1.5 eV 
that gradually increase with $f_{\rm ref}$ [Fig.~\ref{CarbonEigenvaluesNKNK0}(a,b)]. The practical consequences of this observation 
will be discussed in Sec. \ref{ApplicationSection}.

As a conclusion of this preliminary performance evaluation, the NK frozen-orbital correction 
results in a considerable reduction of residual errors $\Pi_{i\sigma}^{u,\rm NK}(f)$,
bringing density-functional approximations in nearly exact agreement with the frozen-orbital linear trend
while exhibiting a slight tendency to underestimate orbital energies. 
In the next sections, we present an extension of the NK correction beyond the frozen-orbital paradigm. 

\subsection{Screened non-Koopmans correction}

\label{ScreenedNKSection}

\begin{figure}[ht!]
\includegraphics[width=8cm]{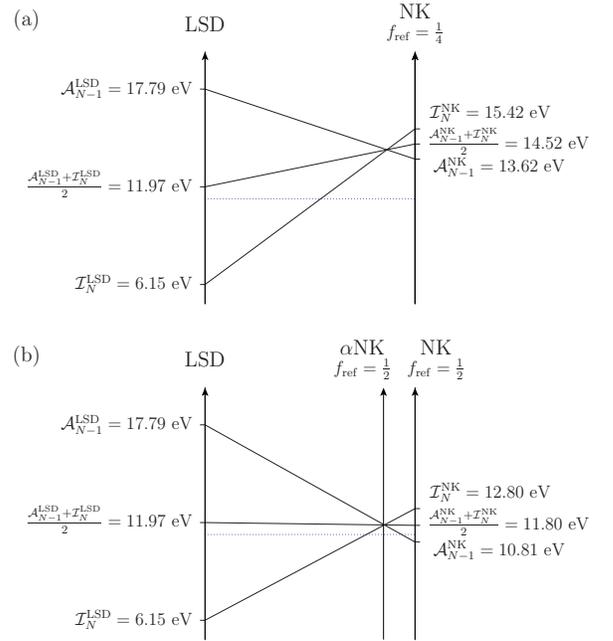}
\caption{LSD differential ionization energies ${\cal A}^{\rm LSD}_{N-1}$,
${\cal I}^{\rm LSD}_N$, and their average $\frac{1}{2}({\cal A}^{\rm LSD}_{N-1}+{\cal I}^{\rm LSD}_N)$ 
compared with NK differential ionization energies 
for (a) $f_{\rm ref}=\frac{1}{4}$ and (b) $f_{\rm ref}=\frac{1}{2}$
for carbon.
\label{NKAdmixture}}
\end{figure}

Having derived and examined the bare non-Koopmans frozen-orbital correction,
we now turn to the analysis of self-interaction for relaxed orbitals.
The effect of the NK correction
on relaxed partial ionization energies for carbon is illustrated in Fig.~\ref{NKAdmixture}. 
The first important observation is that the NK correction decreases
the unphysical convexity of $E_N$, reducing the unphysical discrepancy (${\cal A}_{N-1}-{\cal I}_N$) by a factor
of 6 regardless of the reference occupation $f_{\rm ref}$. 

The second notable feature is 
the fact that the NK correction reverts the convexity trend (i.e., ${\cal I}^{\rm NK}_N>{\cal A}^{\rm NK}_{N-1}$), 
transforming the convex dependence of $E_N$ into a piecewise concave curve.
In fact, bearing in mind that relaxation contributions to $E_N$ are always negative, 
it is relatively straightforward to show that any functional
that satisfies the restricted Koopmans' condition [Eq.~(\ref{RestrictedKoopmansCondition})] is piecewise concave.
Consequently, in contrast to the $\alpha$PZ functional introduced in Sec. \ref{PZSection}, 
there always exists a value of the coefficient $\alpha^{\rm NK}$ ($0\le \alpha^{\rm NK} \le 1$) for which
the $\alpha$NK functional 
\begin{equation}
E^{\alpha \rm NK}=E^{\rm LSD}+\alpha^{\rm NK} (E^{\rm NK}-E^{\rm LSD})
\label{AlphaNKFunctional}
\end{equation}
restores the agreement between ${\cal A}^{\rm NK}_{N-1}$ and ${\cal I}^{\rm NK}_N$.
At this crossing point, the $\alpha$NK relaxed ionization curve is in close agreement with the exact linear trend
described by the generalized Koopmans' condition (see Sec. \ref{AtomicSection}).

Making the approximation that ionization energies vary linearly
with $\alpha^{\rm NK}$, the value of the coefficient for which the crossing 
${\cal A}^{\alpha \rm NK}_{N-1}={\cal I}^{\alpha \rm NK}_N$ occurs can be estimated as
\begin{equation}
\alpha^{\rm NK} \approx \frac{{\cal A}^{\rm LSD}_{N-1}-{\cal I}^{\rm LSD}_N}{({\cal A}^{\rm LSD}_{N-1}
-{\cal I}^{\rm LSD}_N)-({\cal A}^{\rm NK}_{N-1}-{\cal I}^{\rm NK}_N)}.
\label{RelaxationCoefficient}
\end{equation}
This initial estimate can then be refined using the secant-method recursion
\begin{equation}
\alpha_{n+1} = \alpha_n + \frac{(1-\alpha_n)({\cal A}^{\alpha_n \rm NK}_{N-1}-{\cal I}^{\alpha_n \rm NK}_N)}{({\cal A}^{\alpha_n \rm NK}_{N-1}
-{\cal I}^{\alpha_n \rm NK}_N)-({\cal A}^{\rm NK}_{N-1}-{\cal I}^{\rm NK}_N)}.
\label{IterativeRelaxationCoefficient}
\end{equation}
In practice, it is observed that only one or two iterations are sufficient to determine the value of the coefficient
$\alpha^{\rm NK} =  \lim_{n \to \infty} \alpha_n$ and bring the difference
${\cal A}^{\alpha \rm NK}_{N-1}-{\cal I}^{\alpha \rm NK}_N$ below 0.2 eV.

Physically, the coefficient $\alpha^{\rm NK}$ is directly related to the magnitude of orbital relaxation 
upon electron removal.
Indeed, $\alpha^{\rm NK}$ can be viewed as a screening coefficient
whose value is close to 1 for weakly relaxing ionized systems 
(since $E_N^{\rm NK}$ is already piecewise linear in the absence of orbital relaxation)
and is found to be small when relaxation is strong (see Sec. \ref{AtomicSection}).
The corrective factor $\alpha^{\rm NK}$ introduced here is thus endowed with a clear physical interpretation.

As a final note, it should be emphasized that we have adopted here a simple picture of orbital relaxation through
the coefficient $\alpha^{\rm NK}$, which can be viewed as a uniform and isotropic screening factor.
More elaborate screening functions could be employed (at the price of computational complexity).
Such accurate screening approaches provide promising extensions of the $\alpha$NK method and
represent an interesting subject for future studies.

\subsection{Reference transition-state occupation}

\label{ReferenceOccupationSection}


We now proceed to examine the influence of the reference occupation
on the accuracy of calculated total electron removal energies.
To this end, we compare the average of the $\alpha$NK differential electron removal energies 
(that closely approximates the total ionization potential) of carbon 
for $f_{\rm ref}$ different from and equal to $\frac{1}{2}$ in Figs.~\ref{NKAdmixture}(a)
and \ref{NKAdmixture}(b), respectively. 
In the former case, the diagram indicates that the average
\begin{equation}
\frac{1}{2}({\cal A}^{\alpha \rm NK}_{N-1}+{\cal I}^{\alpha \rm NK}_N) \approx  I^{\alpha \rm NK}_N
\end{equation}
deviates significantly from its LSD counterpart,
whereas such deviations do not occur in the case $f_{\text{ref}} = \frac{1}{2}$. 
In fact, from Eqs.~(\ref{IPDefinition}), (\ref{NonKoopmansFunctional2}), and (\ref{AlphaNKFunctional}), 
it can be shown that
\begin{equation}
I^{\alpha \rm NK}_N(f_{\text{ref}}) \approx I^{\rm LSD}_N - \alpha^{\rm NK} \Pi^{u, \rm LSD}_N(f_{\text{ref}})
\end{equation}
(neglecting orbital reconfiguration). Then, substituting the (restricted) Slater's approximation, \cite{Slater1974}
\begin{equation}
\Delta E^{\rm LSD}_{i\sigma} \approx - \epsilon^{u, \rm LSD}_{i\sigma} \left(f_{i\sigma}=\textstyle\frac 12\right)
\label{SlaterApproximationManyElectron}
\end{equation}
into the definition of the non-Koopmans energy contributions [Eq. (\ref{PiEnergy})], one can demonstrate that
\begin{equation}
\Pi^{u,\rm LSD}_N(f_{\text{ref}}=\frac{1}{2}) \approx 0
\end{equation}
to arrive at the relation
\begin{equation}
I^{\alpha \rm NK}_N(f_{\text{ref}}= \frac 12) \approx I^{\rm LSD}_N.
\end{equation}
This result explains the accuracy of the $\alpha$NK correction
in predicting total electron removal energies with the reference occupation $f_{\rm ref}=\frac{1}{2}$. 

It should be noted that in the particular case of one-electron systems, $f_{\rm ref}=\frac{1}{2}$ is not the only possible 
choice for the reference occupation. Indeed, due to the fact the Hartree and exchange-correlation potentials
vanish for empty solitary orbitals, the approximation
\begin{equation}
\Delta E^{\rm LSD} \approx - \epsilon^{u,\rm LSD}(f=0)
\label{SlaterApproximationOneElectron}
\end{equation}
also holds; hence, the solution $f_{\rm ref}=0$ is equally valid. In fact,
$f_{\rm ref}=0$ leads to the exact one-electron Hamiltonian
and the exact solution to the one-electron Schr\"odinger problem.

As an alternative to transition-state arguments, 
the value of $f_{\text{ref}}$ can be justified by inspecting the
sum rule satisfied by the exchange-correlation hole (xc-hole). 
\cite{LangrethPerdew1975,GunnarssonLundqvist1976,PerdewZunger1981}
The xc-hole $h_{\rm xc}$ is defined by the relation
\begin{equation}
E_{\text{xc}} = \frac{1}{2} \int d\mathbf{r} d\mathbf{r}' \frac{\rho(\mathbf{r}) h_{\text{xc}}(\mathbf{r},\mathbf{r}')}
{| \mathbf{r}-\mathbf{r}'|}
\label{xcHoleDefinition}
\end{equation}
and can be explicitly written through the adiabatic connection formalism as
\begin{equation}
\rho(\mathbf{r}) h_{\text{xc}}(\mathbf{r},\mathbf{r}') 
= \big\langle \hat \rho_2(\mathbf{r},\mathbf{r}') \big\rangle_\lambda - \rho(\mathbf{r})\rho(\mathbf{r}')
\label{AdiabaticConnectionA}
\end{equation}
with
\begin{equation}
\big\langle \hat \rho_2(\mathbf{r},\mathbf{r}') \big\rangle_\lambda 
= \int_0^1 d\lambda \langle \Psi_\lambda | \hat \rho_2(\mathbf{r},\mathbf{r}') | \Psi_\lambda \rangle.
\label{AdiabaticConnectionB}
\end{equation}
In Eq.~(\ref{AdiabaticConnectionB}), $\hat \rho_2(\mathbf{r},\mathbf{r}')=\hat{\psi}^{\dagger}(\mathbf{r}) 
\hat{\psi}^{\dagger}(\mathbf{r}') \hat{\psi}(\mathbf{r}') \hat{\psi}(\mathbf{r})$ denotes the pair density operator
and $\Psi_\lambda$ stands for the ground state of a fictitious system where the Coulomb interaction 
is scaled down by $\lambda$ and the effective local single-electron potential $v_{\lambda}$
is added to the Hamiltonian to keep the ground-state density
constant with respect to $\lambda$. 

By definition, \cite{PerdewZunger1981,Gori-GiorgiAngyan2009} the exact xc-hole corresponding to a system with
a fractional number of electrons $N+\omega$, satisfies the sum rule
\begin{multline}
\int d\mathbf{r}' h_{\text{xc}}(\mathbf{r},\mathbf{r}')  =  -1 \\
+\omega (1-\omega) \int_0^1 d\lambda
\frac{\rho_{\lambda,N+1}(\mathbf{r}) -\rho_{\lambda,N}(\mathbf{r})}{\rho(\mathbf{r})},
\label{ExactSumRule}
\end{multline}
where the ground-state density $\rho_{\lambda,M}$ of the system with $M=N$ and $M=N+1$ 
electrons depends on $\lambda$ due to the fact that $M$ differs from $N+\omega$.
It is important to note that the exact xc-hole $h_{\text{xc}}$ integrates to $-1$ only for integer 
electron numbers at variance with the LSD xc-hole $h_{\text{xc}}^{\rm LSD}$, which integrates to $-1$ irrespective of 
the number of electrons.

Turning now to the $\alpha$NK xc-hole $h^{\alpha\text{NK}}_{\text{xc}}$, it is possible to derive 
a relation similar to Eq.~(\ref{ExactSumRule}):
\begin{multline}
\int d\mathbf{r}' h^{\alpha\text{NK}}_{\text{xc}}(\mathbf{r},\mathbf{r}') = -1 \\
+ \alpha^{\rm NK} \sum_{i\sigma} \, f_{i\sigma} ( 2 f_{\text{ref}} -f_{i\sigma} ) 
\frac{n_{i\sigma}(\mathbf{r})}{\rho(\mathbf{r})}
\label{xcHoleSumRuleNK}
\end{multline}
for $N \geq 1$ (the detailed derivation is presented in Appendix \ref{appendix_xc_hole}).
It is then clear that the occupation $f_{\text{ref}}=\frac{1}{2}$ allows to satisfy the xc-hole sum rule
exactly for integer number of electrons, and at least approximately for fractional numbers.
(Note that in the case $N \leq 1 $, the value $f_{\rm ref}=0$ corresponds also to the exact sum-rule.)
For PZ, it has been argued that the enforcement of a similar sum rule
is critical to the quality of the self-interaction correction. \cite{RuzsinszkyPerdew2007}

As a result, the explicit expression of the screened $\alpha$NK functional with $f_{\text{ref}}=\frac{1}{2}$ reads
\begin{widetext}
\begin{multline}
E^{\alpha \rm NK}= E^{\rm LSD} \\ + \alpha^{\rm NK} \sum_{i\sigma} \Bigg[ f_{i\sigma}(1-f_{i\sigma} ) 
E_{\rm H}\left[n_{i\sigma}\right] 
+E^{\rm LSD}_{\rm xc}[\rho-\rho_{i\sigma}]
+ \int d{\bf r} v^{\rm LSD}_{\rm xc,\sigma}\left({\bf r};\left[\rho+(\frac{1}{2}-f_{i\sigma})n_{i\sigma}\right]\right) 
\rho_{i \sigma}({\bf r})
-  E^{\rm LSD}_{\rm xc}[\rho] \Bigg].
\label{HalfOccupationNK}
\end{multline}
\end{widetext}
In the next section, we assess the predictive accuracy of the 
$\alpha$NK functional given by Eq.~(\ref{HalfOccupationNK}), 
first focusing on atoms then extending applications to molecular systems.

\section{Applications}

\label{ApplicationSection}

\subsection{Atomic ionization}

\label{AtomicSection}

\begin{figure*}
\includegraphics[width=15cm]{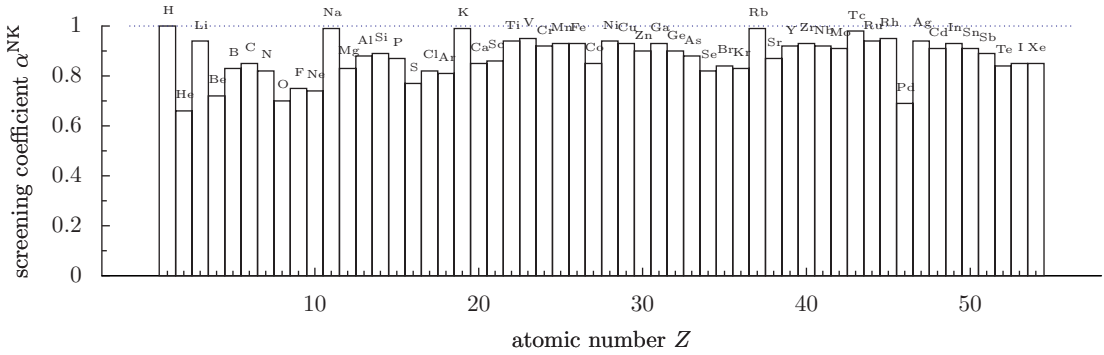}
\caption{Non-Koopmans screening coefficient $\alpha^{\rm NK}$ for the elements of the five first periods. 
\label{AtomicAlpha}}
\end{figure*}

\begin{figure}[ht!]
\includegraphics{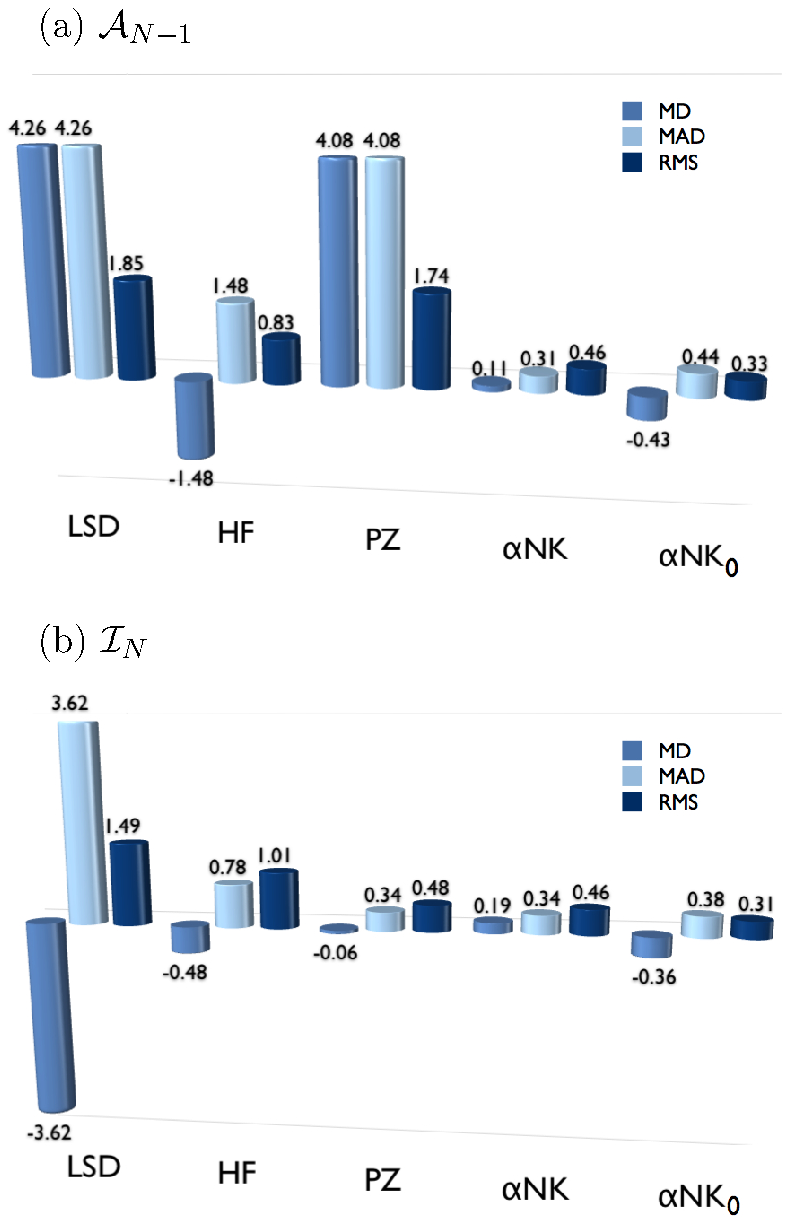}
\caption{{\it Atoms}: mean deviation (MD), mean absolute deviation (MAD), and root mean squared deviation (RMS) of 
LSD, HF, PZ, $\alpha$NK, and $\alpha$NK$_0$ atomic differential electron removal energies (a) ${\cal A}_{N-1}$ 
and (b) ${\cal I}_N$ relative to experiment for the elements of the five first periods. Energy errors are in eV.
\label{AtomicIonizationBars}}
\end{figure}

\begin{figure*}
\includegraphics[width=15cm]{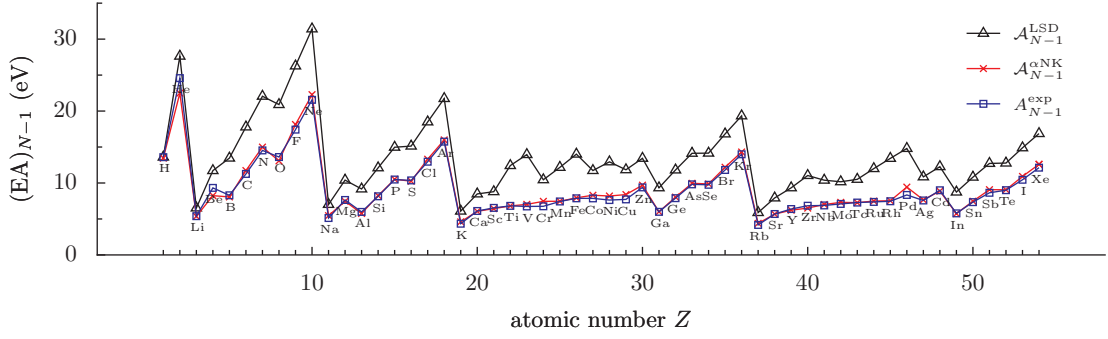}
\caption{LSD and $\alpha$NK differential electron affinities ${\cal A}_{N-1}$ 
(i.e., opposite energy of the lowest unoccupied orbital of the ionized atom X$^+$) 
compared with experiment for the elements of the five first periods. 
\label{AtomicEAs}}
\end{figure*}

\begin{figure*}
\includegraphics[width=15cm]{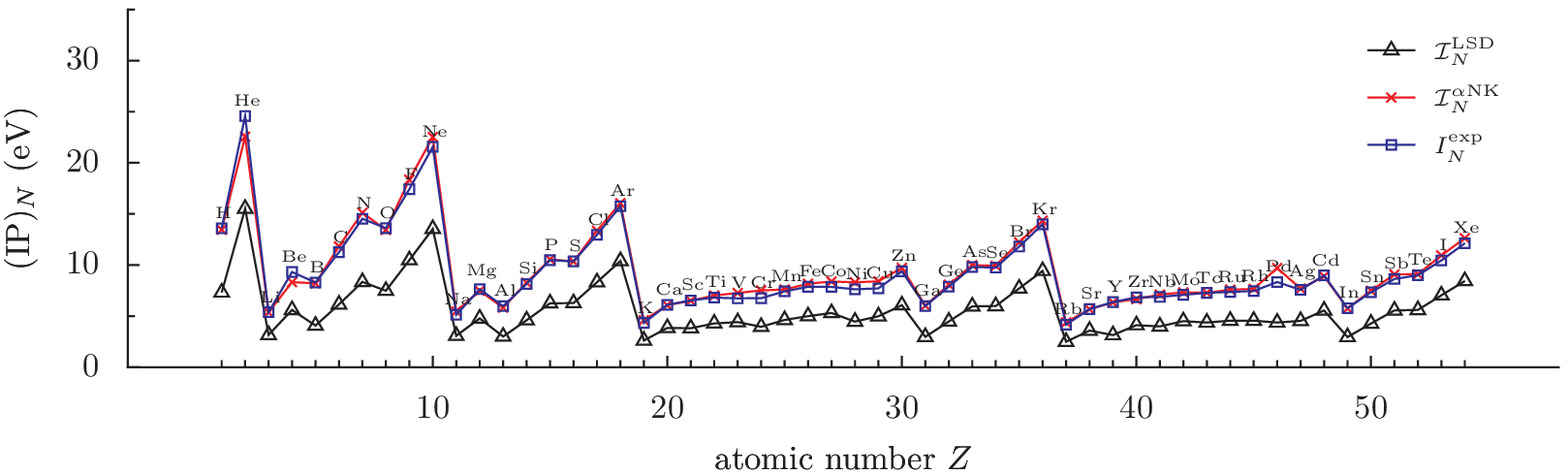}
\caption{LSD and $\alpha$NK differential ionization potentials ${\cal I}_N$ 
(i.e., opposite energy of the highest occupied orbital of the neutral atom X) 
compared with experiment for the elements of the five first periods. 
\label{AtomicIPs}}
\end{figure*}

In order to probe the performance of the $\alpha$NK functional,
we calculate the electron removal energies of a complete range of atomic elements, from hydrogen to xenon,
using the all-electron {\tt LD1} code of the {\sc Quantum-Espresso} distribution. \cite{QuantumEspresso2009} 
(It should be noted that the reported atomic calculations do not account for relativistic effects; the inclusion of scalar-relativistic
contributions results in marginal deviations of $<$80 meV in the absolute precision of $\alpha$NK
differential electron removal energies.) The {\tt LD1} code
proceeds by iterative integration of the spherically symmetric electronic-structure problem
on logarithmic grids. 
The validity of this integration procedure, also employed by Perdew and Zunger in their original work, \cite{PerdewZunger1981} has
been extensively discussed and carefully verified by Goedecker and Umrigar. \cite{GoedeckerUmrigar1997}
More recently, Stengel and Spaldin have shown 
that the spherical approximation is actually required to prevent the appearance of unphysical hybrid states in
the electronic structure of isolated atoms and predict physical
atomic electron removal energies within orbital-dependent self-interaction corrections. \cite{StengelSpaldin2008}

We first compute the screening coefficient 
$\alpha^{\rm NK}$ [Eqs.~(\ref{RelaxationCoefficient}) and (\ref{IterativeRelaxationCoefficient})] for the elements of the five first periodic rows
based on the electronic configurations tabulated in Ref.~[\onlinecite{NISTIonizationTable}] (with the exception of Ni for which
we used the slightly unfavored 4s$^1$3d$^9$ configuration instead of 4s$^2$3d$^8$
that overestimates the experimental ionization potential by more than 1 eV).
The calculated $\alpha^{\rm NK}$ are reported in Fig.~\ref{AtomicAlpha}.
The graph confirms that the screening coefficient varies in a narrow range. Not unexpectedly, the value of $\alpha^{\rm NK}$
is found to be maximal for the elements whose outermost
electronic shell contains a single electron, 
such as hydrogen for which $\alpha^{\rm NK}=1$ and alkali metals, e.g., $\alpha^{\rm NK}({\rm Na})=0.99$. 
In contrast, the coefficient $\alpha^{\rm NK}$ saturates to low values
for filled shell elements, namely, noble gases, e.g., 
$\alpha^{\rm NK}({\rm He})=0.66$, alkaline earth metals, e.g., $\alpha^{\rm NK}({\rm Be})=0.72$, and filled shell transition metals, 
e.g., $\alpha^{\rm NK}({\rm Pd})=0.69$.

After calculating atomic screening coefficients, 
we compare $\alpha$NK differential electron affinity predictions with LSD and experiment in Fig.~\ref{AtomicEAs}.\footnote{
Differential electron affinities can be directly calculated as the opposite energy of the lowest unoccupied orbital
of the ionized atom since the derivative discontinuity contribution defined in Refs.~[\onlinecite{PerdewParr1982}] 
and [\onlinecite{CohenMori-Sanchez2008a}]
is absent for the LSD, HF, PZ, and non-Koopmans functionals.} 
The comparison demonstrates the predictive ability of the $\alpha$NK method,
which brings partial electron removal energies ${\cal A}_{N-1}$ in very close agreement with experimental total electron removal energies
$A_{N-1}^{\rm exp}$, whereas LSD is found to considerably overestimate ${\cal A}_{N-1}$. In quantitative terms,
the differential LSD energy ${\cal A}^{\rm LSD}_{N-1}$ is overestimated by more than 4 eV with a
standard deviation of 1.85 eV [Fig.~\ref{AtomicIonizationBars}(a)]. 
Comparable deviations are obtained with the PZ self-interaction correction. The HF energy
${\cal A}^{\rm HF}_{N-1}$ are instead underestimated by a smaller margin of 1.48 eV. 
The $\alpha$NK correction results in substantial improvement 
in the calculation of partial electron removal energies, reducing the error to 0.31 eV. 
Here, it is quite interesting to note that the $\alpha$NK variational contributions counterbalance the slight
tendency of the $\alpha$NK$_0$ correction to underestimate electron removal energies within LSD.

We now examine partial ionization potential predictions (Fig.~\ref{AtomicIPs}).
A marked difference with the above electron affinity results is the enhanced accuracy of the HF and PZ theories. 
The improved performance in predicting atomic ionization potentials results from the fact that orbital relaxation 
compensates the absence of correlation contributions in HF and
cancels residual non-Koopmans errors in PZ. \cite{PerdewZunger1981}
Nevertheless, even with beneficial error cancellation in favor of HF and PZ, the $\alpha$NK deviation
is still the lowest, approximately equal to the PZ mean absolute error of 0.34 eV [Fig.~\ref{AtomicIonizationBars}(b)]. 

The precision of $\alpha$NK differential ionization energies reflects the intrinsic accuracy of the
underlying LSD total energy functional in reproducing subtle atomic ionization trends
that are difficult to describe with, e.g., semi-empirical approximations. \cite{PucciMarch1982}

\begin{figure}[ht!]
\includegraphics[width=8.5cm]{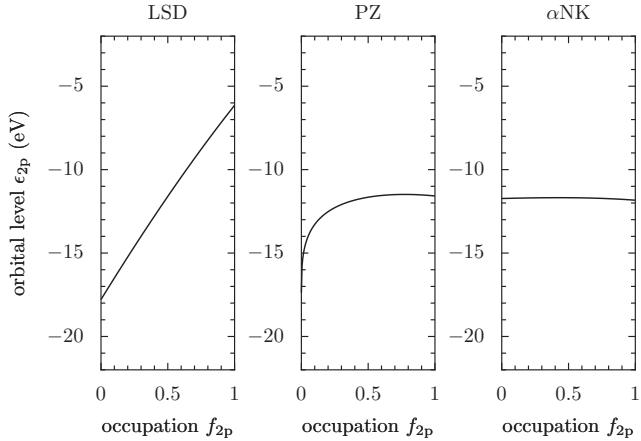}
\caption{LSD, PZ, and $\alpha$NK relaxed orbital energies of the highest occupied state of carbon.
\label{CarbonRelaxedEigenvalues}}
\end{figure}

To complement these observations, the full dependence of the LSD, PZ, and $\alpha$NK relaxed energies
$\epsilon_{\rm 2p}$ of the highest occupied atomic orbital of carbon as a function of its occupation
is depicted in Fig.~\ref{CarbonRelaxedEigenvalues}.
Confronting the LSD and PZ graphs with those presented in Fig.~\ref{CarbonEigenvalues},
it is seen that orbital relaxation causes a non-negligible decrease of the unphysical shift $\epsilon_{\rm 2p}(1)-\epsilon_{\rm 2p}(0)$ 
of 1.5 eV for both functionals.
Additionally, the PZ orbital energy $\epsilon^{\rm PZ}_{\rm 2p}$ becomes less curved at higher occupations, 
confirming that orbital relaxation enhances the performance of the PZ correction. \cite{PerdewZunger1981} 
Nevertheless, the inflexion of the curve remains important in the vicinity of $f_{\rm 2p}=0$
due to the fact that the PZ correction leaves the energy of the empty state unchanged. We observe that this unphysical trend
is almost completely removed by the $\alpha$NK correction, clearly showing that the screened $\alpha$NK method 
is apt at imposing the generalized Koopmans' condition for any fractional value of $f_{\rm 2p}$.

The above comparisons demonstrate the predictive performance of the non-Koopmans method in correcting
atomic differential electron affinities and first ionization potentials, placing
${\cal A}_{N-1}$ and ${\cal I}_N$ predictions on the same level of accuracy with respect to experiment. The fact that $\alpha$NK
improves ${\cal A}_{N-1}$ and ${\cal I}_N$ with the same precision ensures
the accuracy of non-Koopmans total energy differences and related equilibrium properties. 
The results presented in the next sections provide further support to this conclusion.

\subsection{Molecular ionization}

\label{MolecularSection}

\begin{figure}[ht!]
\includegraphics[width=8cm]{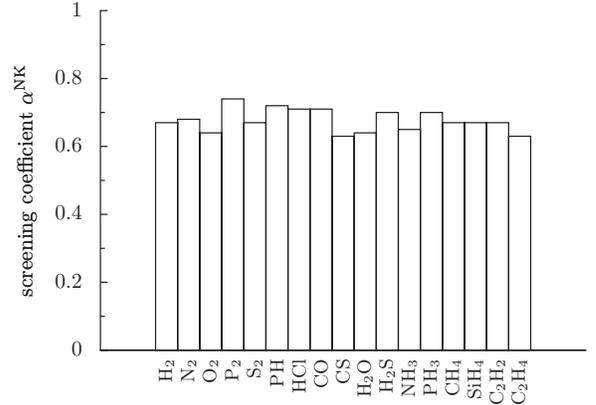}
\caption{Molecular non-Koopmans screening coefficient $\alpha^{\rm NK}$. 
\label{MolecularAlpha}}
\end{figure}

\begin{table*}
\caption{LSD, HF, PZ, $\alpha$NK, and $\alpha$NK$_0$ differential molecular electron removal energies compared 
with experimental vertical electron removal energies.
Mean deviations (MD), mean absolute deviations (MAD), and root mean squared deviations of the error (RMS) in
absolute and relative terms are also reported.
The adiabatic ionization lower bound is given when the experimental vertical ionization
energy is not available. Energies are in eV.
\label{MolecularTable}}
\begin{tabular*}{\textwidth}{@{\extracolsep{\fill}}lccccccccccc}
\hline \hline 
&&&&&&&&&&&\\
 & \multicolumn{2}{c}{LSD} & \multicolumn{2}{c}{HF} & \multicolumn{2}{c}{PZ} & \multicolumn{2}{c}{$\alpha$NK} & \multicolumn{2}{c}{$\alpha$NK$_0$} & 
Exp.\footnotemark[1]\\
     & ${\cal A}_{N-1}$ & ${\cal I}_{N}$ & ${\cal A}_{N-1}$ & ${\cal I}_{N}$ & ${\cal A}_{N-1}$ & ${\cal I}_{N}$ & ${\cal A}_{N-1}$ & ${\cal I}_{N}$ 
& ${\cal A}_{N-1}$ & ${\cal I}_{N}$ & \\
\\
\hline 
\\
H$_2$      & 18.84 & 10.16  & 14.78  & 16.24 & 19.01 & 16.99 & 14.79 & 14.83 & 15.74 & 15.69 & $>$15.43\\
N$_2$      & 20.85 & 10.37  & 12.77  & 17.15 & 21.55 & 17.78 & 16.14 & 16.20 & 15.38 & 15.52 & $>$15.58\\
O$_2$      & 18.94 &  7.20  &  9.27  & 14.71 & 18.11 & 15.43 & 13.85 & 13.99 & 13.04 & 12.51 & 12.30 \\
P$_2$      & 14.28 &  7.26  &  8.56  & 10.84 & 14.40 & 11.53 & 10.86 & 10.85 & 10.35 & 10.35 & 10.62 \\
S$_2$      & 13.28 &  5.81  &  7.49  & 10.49 & 13.05 & 11.06 & 10.17 & 10.19 &  9.59 &  9.54 &  9.55 \\
PH         & 13.92 &  5.81  &  8.39  & 10.25 & 13.81 & 10.84 &  9.94 & 10.02 &  9.68 &  9.68 & $>$10.15\\
HCl        & 17.81 &  8.11  & 10.33  & 13.05 & 17.58 & 13.82 & 13.09 & 13.17 & 12.30 & 12.32 & $>$12.75\\       
CO         & 18.70 &  9.14  & 11.03  & 15.06 & 18.45 & 15.40 & 14.15 & 14.24 & 13.68 & 13.74 & 14.01 \\
CS         & 15.20 &  7.42  &  7.30  & 12.83 & 14.53 & 13.29 & 11.63 & 11.76 & 10.83 & 10.89 & $>$11.33\\
H$_2$O     & 18.97 &  7.33  &  8.97  & 13.81 & 18.50 & 14.77 & 13.17 & 13.45 & 11.75 & 11.94 & $>$12.62\\
H$_2$S     & 14.87 &  6.39  &  8.17  & 10.55 & 14.70 & 11.52 & 10.77 & 10.86 & 10.15 & 10.17 & 10.50 \\
NH$_3$     & 16.07 &  6.23  &  7.61  & 11.55 & 15.75 & 12.50 & 11.15 & 11.40 & 10.18 & 10.31 & 10.82 \\
PH$_3$     & 14.50 &  6.84  &  8.29  & 10.46 & 14.15 & 11.45 & 10.77 & 10.83 & 10.41 & 10.41 & 10.59 \\
CH$_4$     & 18.69 &  9.46  & 11.95  & 14.92 & 18.98 & 16.20 & 14.46 & 14.51 & 13.91 & 13.86 & 13.60 \\
SiH$_4$    & 15.86 &  8.50  & 10.84  & 13.26 & 16.46 & 14.35 & 12.67 & 12.68 & 12.53 & 12.40 & 12.30 \\
C$_2$H$_2$ & 16.19 &  7.40  &  8.79  & 11.40 & 16.19 & 12.97 & 12.05 & 12.14 & 11.11 & 11.15 & 11.49 \\
C$_2$H$_4$ & 15.12 &  7.01  &  7.91  & 10.43 & 15.07 & 12.62 & 11.43 & 11.53 & 10.50 & 10.54 & 10.68 \\
\\
MD         & 4.57  & --4.35 & --2.46 & 0.75  &  4.47 &  1.66 &  0.40 & 0.49  &--0.19 &--0.19 & --- \\
           & 38.5\%&--36.4\%&--21.0\%&  5.9\%& 37.4\%& 13.7\%&  3.4\%&  4.2\%&--1.7\%&--1.7\%& --- \\
MAD        & 4.57  & 4.35   &  2.46  & 0.80  & 4.47  & 1.66  & 0.50  & 0.58  & 0.38  & 0.29  & --- \\ 
           & 38.5\%& 36.4\% &  21.0\%& 6.4\% & 37.4\%& 13.7\%& 4.1\% & 4.8\% & 3.2\% & 2.5\% & --- \\
RMS        & 0.95  & 0.63   &  0.83  & 0.73  & 0.89  & 0.66  & 0.46  & 0.48  & 0.40  & 0.28  & --- \\
           & 7.7\% & 4.0\%  &  7.4\%& 5.9\%  & 6.3\% & 5.0\% & 3.7\% & 3.9\% & 3.3\% & 2.4\% & --- \\
\\
\hline \hline
\end{tabular*}
\flushleft
\footnotemark[1]{Reference [\onlinecite{NISTMolecularIonization}].}
\end{table*}

In this section, we focus on the study of molecular systems. For this purpose, we have implemented 
the HF, PZ, and $\alpha$NK methods in the plane-wave pseudopotential {\tt CP} (Car-Parrinello) code of the
{\sc Quantum-Espresso} distribution. \cite{QuantumEspresso2009} In this code, orbital
optimization proceeds via fictitious Newtonian damped electronic dynamics.

The main difficulty in the {\tt CP} implementation of the HF, PZ, and $\alpha$NK functionals
is the correction of periodic-image errors that arise from the use of the
supercell approximation. \cite{PayneTeter1992} Such numerical errors preclude the accurate evaluation
of exchange terms and orbital electrostatic potentials. 
To eliminate periodic-image errors in the plane-wave evaluation of 
exchange and electrostatic two-electron integrals, we employ
countercharge correction techniques. \cite{DaboKozinsky2008}
In addition to this difficulty, explicit orthogonality constraints must be considered for the accurate calculation 
of the gradient of the orbital-dependent PZ and $\alpha$NK functionals. \cite{GoedeckerUmrigar1997} To incorporate these additional 
constraints, we use the efficient iterative orthogonalization cycle implemented in the original {\tt CP}
code. \cite{LaasonenPasquarello1993}
In terms of computational performance, the cost of $\alpha$NK calculations is here only 40\% higher than that of
PZ and lower than that of HF.

In Table~\ref{MolecularTable}, we compare LSD, HF, PZ, and NK partial electron removal energy predictions for
a representative set of molecules. In each case, molecular geometries are fully relaxed (the accuracy 
of equilibrium geometry predictions will be examined in Sec. \ref{GeometrySection}). To perform our calculations, 
we employ LSD norm-conserving pseudopotentials \cite{Pickett1989} with an energy cutoff of 60 Ry for the plane-wave 
expansion of the electronic wavefunctions. With this calculation parameter, 
we verify that ${\cal A}_{N-1}$ and ${\cal I}_{N}$ are converged to within less than 50 meV. 

It is frequently argued that substituting LSD pseudopotentials for their HF, PZ, and NK counterparts
has minor effect on the predicted energy differences. \cite{GoedeckerUmrigar1997} Comparing our
pseudopotential calculations with all-electron atomic results (see Sec. \ref{AtomicSection}), we actually found
that the use of LSD pseudopotentials yields HF, PZ, and NK
electron removal energies with a typical error of 0.1 to 0.2 eV. However,
since these moderate deviations affect HF, PZ, and NK predictions in identical manner,
the pseudopotential substitution does not alter the validity of the present comparative analysis. 

As expected, one conspicuous feature in Table~\ref{MolecularTable} is the poor performance of LSD
that predicts molecular partial electron removal energies with an average error of $\pm$40\%. As was the case for atoms,
the PZ self-interaction correction reduces the error in predicting  ${\cal I}_N$ to less than 14\%, which corresponds
to an average deviation of 1.68 eV, whereas ${\cal A}_{N-1}$ predictions are not improved. In comparison,
$\alpha$NK partial ionization energies are predicted with a remarkable precision of 0.50 eV (4.1\%) 
and 0.58 eV (4.8\%) for ${\cal A}_{N-1}$ and ${\cal I}_N$, respectively. The $\alpha$NK accuracy
in predicting molecular vertical ionization energies compares favorably (arguably, even more accurately) with that of recently published
fully self-consistent GW many-body perturbation theory calculations. \cite{RostgaardJacobsen2010}

It should also be noted that we perform here only 
one iteration [Eq.~(\ref{RelaxationCoefficient})] to determine the screening coefficient and that
the calculated $\alpha^{\rm NK}$ vary in a very limited range of values, 
even narrower than that found in the case of atoms (Fig.~\ref{MolecularAlpha}).

To conclude the analysis of ${\cal A}_{N-1}$ and ${\cal I}_N$ predictions, 
we focus on the influence of the variational contributions 
$\hat w^{\rm LSD}_{{\rm ref},i \sigma}$ and $\hat w^{\rm LSD}_{{\rm xd},i \sigma}$
on the accuracy of the $\alpha$NK total energy method.
Similarly to the comparative analysis presented in Sec. \ref{ComparativeAssessmentSection},
we use the $\alpha$NK$_0$ orbital-energy formulation to evaluate the magnitude of 
$\hat w^{\rm LSD}_{{\rm ref},i \sigma}$ and $\hat w^{\rm LSD}_{{\rm xd},i \sigma}$ errors.
In agreement with the trend already observed, 
$\alpha$NK$_0$ predictions for ${\cal A}_{N-1}$ and ${\cal I}_N$ are lower than $\alpha$NK electron removal energies with
negative shifts of 0.59 eV and 0.68 eV, respectively.
(Thus, $\alpha$NK$_0$ results are found to be even closer to experiment than their $\alpha$NK 
counterparts with mean absolute error margins of only 3.2\% for  ${\cal A}_{N-1}$ and 2.5\%  for ${\cal I}_N$.)
This direct comparison indicates that
$\hat w^{\rm LSD}_{{\rm ref},i \sigma}$ and $\hat w^{\rm LSD}_{{\rm xd},i \sigma}$ introduce
non-negligible energy shifts in the calculation of frontier orbital levels. However,
these errors are much smaller than typical self-interaction deviations of 4 to 5 eV,
providing quantitative justifications of the excellent precision of the $\alpha$NK method.

\subsection{Equilibrium structural properties}

\label{GeometrySection}

\begin{figure}[ht!]
\includegraphics{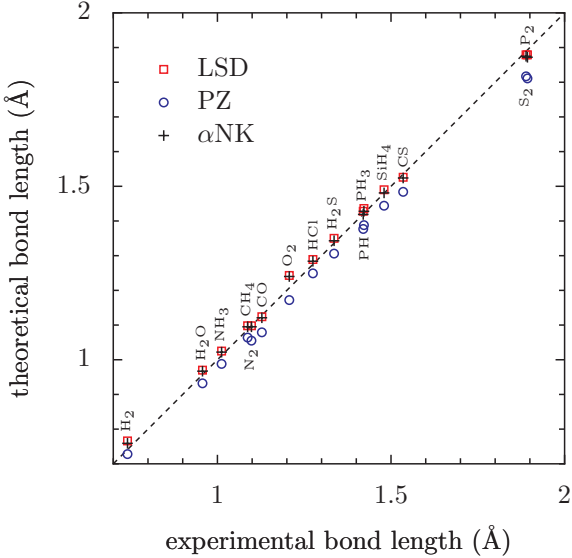}
\caption{LSD, PZ, and $\alpha$NK molecular bond lengths compared with experiment (Ref.~[\onlinecite{CRC2009}]).
\label{MolecularGeometries}}
\end{figure}

\begin{figure}[ht!]
\includegraphics{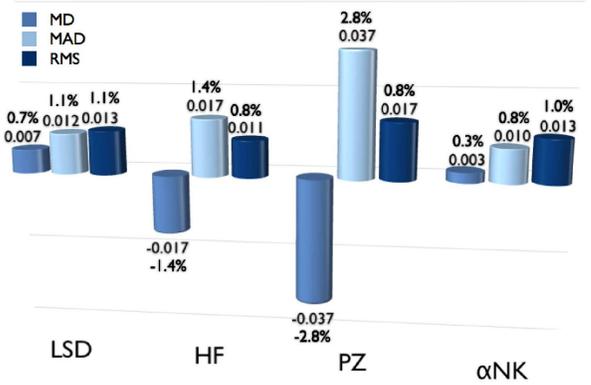}
\caption{{\it Molecular geometries}: LSD, HF, PZ, and $\alpha$NK mean
deviations (MD), mean absolute deviations (MAD), and root mean squared deviations of the error (RMS)
with respect to experiment of the predicted bond lengths for the 17 molecules listed in Table~\ref{MolecularTable}.
Error bars are in \AA. Relative errors are also reported.
\label{MolecularBars}}
\end{figure}

After analyzing partial electron removal energies, we now report on 
the accuracy of $\alpha$NK equilibrium geometry calculations. 
We compare LSD, PZ, and $\alpha$NK structural predictions to experimental bond lengths in Fig.~\ref{MolecularGeometries}
and we present LSD, HF, PZ, and $\alpha$NK error bars in Fig.~\ref{MolecularBars}.
The first important observation is the very good accuracy of LSD predictions
with a mean absolute relative error of 1.1\% for the seventeen molecules listed in Table~\ref{MolecularTable}.
PZ bond lengths are instead sensibly underestimated with a mean uncertainty of 2.8\%. 
In contrast with PZ calculations, $\alpha$NK results deviate from experiment by a relative error margin of
0.8\%, which is lower than that of LSD, demonstrating that the $\alpha$NK self-interaction correction does not 
deteriorate and even improves LSD structural predictions, at variance with the conventional PZ self-interaction correction.

These results illustrate the tendency of PZ to overbind molecular structures,
and confirm the systematic improvement brought about by the $\alpha$NK correction. The promising potential of
the $\alpha$NK correction in predicting other thermodynamical properties (e.g., dissociation energies and vibrational
frequencies) will be critically explored in a separate study.

\subsection{Photoemission energies}

\begin{table}
\caption{LSD, HF, PZ, and $\alpha$NK orbital energies of neon, argon, and krypton compared with 
experimental photoemission energies. 
Relative mean absolute deviations (MAD) with respect to experimental photoionization results are also reported.
The experimental photoemission energies of the spin-orbit doublets
of p and d orbitals are indicated. Computational photoionization predictions 
do not include spin-orbit coupling. Energies are in eV. 
\label{AtomicPhotoemission}}
\begin{tabular*}{0.475\textwidth}{@{\extracolsep{\fill}}lcrrrrr}
\hline \hline
\\
    &    &    LSD &    HF &    PZ & $\alpha$NK & Exp.\footnotemark[1] \\
\\
\hline \\
Ne    &  2p  &   13.54 &   23.11 &   22.91  &   22.52 &    21.6--21.7\\
      &  2s  &   35.99 &   52.49 &   45.13  &   45.11 &    48.5\\
      &  1s  &  824.68 &  891.75 &  889.41  &  872.14 &   870.2\\
\\                  
Ar    &  3p  &   10.40 &   16.05 &   15.76  &   16.04 &    15.7--15.9\\
      &  3s  &   24.03 &   34.74 &   30.22  &   30.54 &    29.3\\
      &  2p  &  229.77 &  260.45 &  256.12  &  254.65 &   248.4--250.6\\
      &  2s  &  293.73 &  335.30 &  315.49  &  315.40 &   326.3\\
      &  1s  & 3096.69 & 3227.47 & 3218.88  & 3193.55 &  3205.9\\
\\                  
Kr    &  4p  &    9.43 &   14.25 &   13.97  &   14.35 &    14.1--14.2\\
      &  4s  &   22.33 &   31.34 &   27.78  &   28.27 &    27.5\\
      &  3d  &   83.65 &  104.06 &  101.29  &  101.67 &    93.8--95.0\\
      &  3p  &  192.84 &  226.70 &  209.04  &  210.71 &   214.4--222.2\\
      &  3s  &  253.48 &  295.21 &  269.48  &  271.24 &   292.8\\
      &  2p  & 1633.17 & 1714.53 & 1695.09  & 1692.46 &  1678.4--1730.9\\
      &  2s  & 1803.75 & 1902.11 & 1852.00  & 1853.32 &  1921.0\\
      &  1s  &13877.37 &14154.31 &14128.17  &14080.07 & 14326.0\\
\\
\multicolumn{2}{l}{MAD}       & 19.2\% &  4.5\%   & 3.3\%    &  3.2\%  & ---    \\ 
\\
\hline \hline
\end{tabular*}
\flushleft
\footnotemark[1]{Reference [\onlinecite{CRC2009}].}
\end{table}

\begin{table}
\caption{LSD, HF, PZ, and $\alpha$NK$_0$ orbital energies of benzene compared with 
experimental photoemission energies. 
Relative mean absolute deviations (MAD) with respect to experimental photoionization results are also reported.
Energies are in eV. 
\label{BenzenePhotoemission}}
\begin{tabular*}{0.475\textwidth}{@{\extracolsep{\fill}}lcccccc}
\hline \hline
\\
        &    LSD &    HF &    PZ & $\alpha$NK & $\alpha$NK$_0$ & Exp.\footnotemark[1] \\
\\
\hline \\
e$_{1g}$ &    6.59 &  9.18 &  9.43 & 10.39 &  9.39 &  9.3  \\                 
e$_{2g}$ &    8.28 & 13.54 & 15.46 & 12.66 & 12.48 & 11.8  \\                 
a$_{2u}$ &    9.43 & 13.64 & 12.99 & 13.25 & 12.60 & 12.5  \\                 
e$_{1u}$ &   10.33 & 16.02 & 17.67 & 14.75 & 14.56 & 14.0  \\                 
b$_{2u}$ &   11.02 & 16.95 & 18.40 & 15.46 & 15.15 & 14.9  \\                 
b$_{1u}$ &   11.26 & 17.51 & 18.82 & 15.65 & 15.69 & 15.5  \\                 
a$_{1g}$ &   13.10 & 19.26 & 20.60 & 17.58 & 17.30 & 17.0  \\                 
e$_{2g}$ &   14.85 & 22.39 & 22.72 & 19.27 & 19.34 & 19.2  \\
\\
MAD      &  26.1\% & 12.0\%& 18.4\%& 4.9\% & 2.1\% & --- \\
\\
\hline \hline
\end{tabular*}
\flushleft
\footnotemark[1]{Reference [\onlinecite{PiancastelliKelly1987}].}
\end{table}

\begin{table*}
\caption{LSD, HF, PZ, and $\alpha$NK orbital energies of fullerene C$_{60}$ compared with 
constrained LSD total energy differences ($\Delta_{\rm c}$LSD) and experimental photoemission energy bands. Energies are in eV.
\label{FullerenePhotoemission}}
\begin{tabular*}{\textwidth}{@{\extracolsep{\fill}}cccccccccccc}
\hline \hline
\\
band &    
\multicolumn{2}{c}{LSD} & 
\multicolumn{2}{c}{HF}  & 
\multicolumn{2}{c}{PZ}  & 
\multicolumn{2}{c}{$\alpha$NK}  &       
\multicolumn{2}{c}{$\Delta_{\rm c}$LSD\footnotemark[1]} & Exp.\footnotemark[2]      \\
\\
\hline \\
\scriptsize I   & h$_u$ &  5.84 & h$_u$ &  7.49 & h$_u$ &  8.77 & h$_u$ & 7.45  & h$_u$ & 7.61  & 7.60  \\                 
\\                                                        
\scriptsize II  & g$_g$ &  7.03 & h$_g$ &  9.42 & g$_g$ &  9.80 & g$_g$ & 8.64  & g$_g$ & 8.78  & 8.95  \\                 
                & h$_g$ &  7.15 & g$_g$ &  9.64 & h$_g$ & 10.48 & h$_g$ & 8.75  & h$_g$ & 8.90  &       \\                 
\\                                                        
\scriptsize III & h$_u$ &  8.72 & g$_u$ & 12.42 & g$_u$ & 12.21 & h$_u$ & 10.31 & h$_u$ & 10.47 & 10.82--11.59  \\                 
                & g$_u$ &  8.74 & t$_u$ & 12.99 & t$_u$ & 12.69 & g$_u$ & 10.35 & g$_u$ & 10.50 &  \\                 
                & h$_g$ &  9.03 & h$_u$ & 13.08 &       &       & h$_g$ & 10.64 & h$_g$ & 10.79 &  \\                 
                & t$_u$ &  9.28 &       &       &       &       & t$_u$ & 10.91 & t$_u$ & 11.03 &  \\                 
\\                                                        
\scriptsize IV  & g$_u$ & 10.05 & h$_g$ & 13.46 & h$_u$ & 14.13 & g$_u$ & 11.66 & g$_u$ & 11.79 & 12.43--13.82 \\                 
                & t$_g$ & 10.52 & g$_u$ & 15.06 & t$_g$ & 15.41 & t$_g$ & 12.12 & t$_g$ & 12.28 &    \\
                & h$_g$ & 10.59 & t$_g$ & 15.20 & h$_g$ & 15.81 & h$_g$ & 12.20 & h$_g$ & 12.33 &    \\
                &       &       & h$_g$ & 15.66 &       &       &       &       &       &       &    \\
\\
\hline \hline
\end{tabular*}
\flushleft
\footnotemark[1]{Reference [\onlinecite{TiagoKent2008}].} \\
\footnotemark[2]{Reference [\onlinecite{LichtenbergerNebesny1991}].}
\end{table*}

Having validated the non-Koopmans self-interaction correction for the calculation of electron removal energies
and equilibrium structures, we now evaluate
the performance of the $\alpha$NK and $\alpha$NK$_0$ 
methods in predicting photoemission energies,
for which LSD and GGAs exhibit notable failures.

From the theoretical point of view, the very poor performance of LSD and GGA is expected;
Kohn-Sham density-functional theory eigenvalues are not meant to predict
excited-state properties.~\cite{Onida2002,Gatti2007} 
In practice, total electron removal energies computed from constrained density-functional calculations ($\Delta_{\rm c}$SCF)
are typically found to be in good agreement with experiment. \cite{SchipperGritsenko2000,TiagoKent2008}
This level of accuracy suggests that orbital-dependent self-interaction corrected functionals can provide orbital energies in accordance with
spectroscopic results.\cite{ChongGritsenko2002} 
This expectation is confirmed by PZ photoemission predictions for neon, argon, and krypton that
we reproduce here using the {\tt LD1} code (Table~\ref{AtomicPhotoemission}).

Nevertheless, similarly to the trend observed in Sec. \ref{MolecularSection},
the predictive ability of PZ deteriorates in the case of molecular photoionization; this is 
at variance with the $\alpha$NK method. 
To illustrate this fact, we compare LSD, HF, PZ, and non-Koopmans predictions
for the photoemission spectrum (PES) of benzene in Table~\ref{BenzenePhotoemission}
and fullerene in Table~\ref{FullerenePhotoemission}.
In the molecular photoemission calculations, we use the {\tt CP} code with the computational procedure
described in Sec.~\ref{MolecularSection}. We employ fully relaxed geometries for benzene
and the LSD atomic structure of C$_{60}$, which is found to be in excellent agreement with the NMR experimental
geometry. \cite{FeustonAndreoni1991} 

Focusing first on benzene, we observe that LSD underestimates electron binding energies with errors as large as 
4.35 eV for low-lying states. In contrast to LSD, the HF theory provides overestimated photoemission energies
with absolute deviations that increase gradually from 0.12 to 3.19 eV when approaching the bottom of the PES.
Similar trends are observed for PZ with the difference that the errors do not systematically increase
with increasing photoemission energies, leading in particular to the
incorrect ordering of the e$_{2g}$ and a$_{2u}$ levels. In contrast, $\alpha$NK 
restores the correct relative peak positions and
yields slightly overestimated electron binding energies with an absolute precision of $4.9$\%. 
The slight tendency of $\alpha$NK to overestimate electron
binding energies is here again due to the influence of variational contributions,
as directly confirmed by the performance of the $\alpha$NK$_0$ orbital-energy method,
which predicts photoemission energies in remarkable agreement with experiment. 

Similarly to benzene, LSD energy predictions for fullerene are significantly 
underestimated. However, since the dispersion of the errors is much narrower
than in the case of benzene, a simple shift 
of LSD photoemission bands, equal to the difference between the theoretical and experimental 
HOMO levels, can bring the predicted PES in close agreement with experiment. \cite{FeustonAndreoni1991}
Despite the excellent precision of HF in the top region of the spectrum, HF photoemission energies
are largely overestimated for low-lying states. 
In addition, HF inverts the h$_g$ and g$_g$ states in the second photoemission band although
it predicts the correct peak ordering in the third and fourth bands. \cite{TiagoKent2008} 
The performance of PZ is found to be slightly worse than that of HF with 
significant qualitative errors in the grouping and ordering of the states. 
In contrast, $\alpha$NK correctly shifts the
spectrum and brings photoemission energies in very good agreement with experiment.
Predicted $\alpha$NK binding energies are also in excellent agreement with constrained LSD total energy differences, \cite{TiagoKent2008}
providing a final validation of the performance of the $\alpha$NK self-interaction correction 
in bringing physical meaning to orbital energies --- i.e., in identifying orbital energies as opposite total electron removal energies.

\section{Summary and outlook}

In summary, we have demonstrated that the correction of the nonlinearity of the ground-state energy $E_N$
as a function of the number of electrons $N$, at the origin of 
important discrepancies between total and differential electron removal energies, 
and related fundamental qualitative and quantitative self-interaction errors, can be achieved without altering the 
otherwise excellent performance of density-functional approximations in describing systems with non-fractional
occupations. To construct the non-Koopmans self-interaction correction,
we have first defined an exact non-Koopmans measure of self-interaction and
adopted the frozen-orbital approximation (i.e., the framework of the restricted Koopmans' theorem) 
as a working alternative to the conventional one-electron paradigm. We have then accounted for orbital relaxation
by introducing the screening coefficient $\alpha^{\rm NK}$, which bears the physical significance
of a uniform and isotropic screening factor that can be determined iteratively --- thereby closely satisfying the generalized 
Koopmans' condition. This self-interaction correction scheme
can be applied to any local, semilocal or hybrid density-functional approximation.
The remarkable predictive performance of the non-Koopmans theory 
has been demonstrated for a range of atomic and molecular systems.

The theory developed here represents a significant step in the correction of 
electron self-interaction in electronic-structure theories. 
Nevertheless, interesting problems are left open. One central question is that
similarly to the PZ approach, the $\alpha$NK method leads to an orbital-dependent Hamiltonian, although
it is always possible in principle to derive a consistent density-dependent formulation using, e.g., 
optimized effective potential mappings. \cite{KummelKronik2008}
It is a long held tenet that the orbital dependence of self-interaction functionals and the subsequent loss of invariance with
respect to unitary transformation of the one-body density matrix precludes applications
to periodic systems (e.g., conjugate polymers and crystalline materials). 
In future studies, we will explore solutions to this central conceptual
difficulty without resorting to density-dependent unitary invariant mappings.

\acknowledgments

The authors are indebted to X. Qian, N. Laachi, S. de Gironcoli, \'E. Canc\`es, and
T. K\"orzd\"orfer for helpful suggestions and fruitful comments.
The computations in this work have been performed using the 
{\sc Quantum-Espresso} package (http://www.quantum-espresso.org) \cite{QuantumEspresso2009} 
and computational resources offered by the Minnesota Supercomputing Institute 
and DE-FG02-05ER46253.

I. D. and N. M. acknowledge support from MURI grant DAAD 19-03-1-016.
I. D., Y. L., and M. C. acknowledge support from the Grant in Aid of the University 
of Minnesota and from grant ANR 06-CIS6-014.  M. C. acknowledges partial support 
from NSF grant EAR-0810272 and 
from the Abu Dhabi-Minnesota Institute for Research Excellence (ADMIRE).
N. M., A. F., and N. P. acknowledge support from DOE SciDAC DE-FC02-06ER25794,
DE-FG02-05ER15728, and MIT-ISN.

\appendix


\begin{widetext}

\section{Screened non-Koopmans exchange-correlation hole sum rule}

\label{appendix_xc_hole}

In this appendix, we derive the explicit expression of the $\alpha$NK xc-hole.
Starting from the relation
\begin{equation}
      E^{\alpha\text{NK}} = E^{\text{LSD}} + \alpha^{\text{NK}} \sum_{i\sigma} \Big(
                  E^{\text{LSD}}_{\text{Hxc}}[\rho -\rho_{i\sigma}] -
                  E^{\text{LSD}}_{\text{Hxc}}[\rho]
                  +\int d\mathbf{r} \rho_{i\sigma}(\mathbf{r})
                  v_{\text{Hxc},\sigma}^{\text{LSD}}(\mathbf{r};[\rho_{i\sigma}^{\text{ref}}])
                  \Big)
\label{NKFunctional}
\end{equation}
and from the definition of the xc-hole [Eq.~(\ref{xcHoleDefinition})], the contributions 
to the total $\alpha$NK xc-hole arising from the three summation terms in Eq.~(\ref{NKFunctional})
can be worked out.
Those terms will be labeled $h_{\text{xc},i\sigma}^{\rm (a)}$, $h_{\text{xc},i\sigma}^{\rm (b)}$, and 
$h_{\text{xc},i\sigma}^{\rm (c)}$, respectively.
Focusing on the first term, it is straightforward to obtain
\begin{equation}
h_{\text{xc},i\sigma}^{\rm (a)}(\mathbf{r},\mathbf{r}') =  
\frac{\rho(\mathbf{r}) -\rho_{i\sigma}(\mathbf{r})}{\rho(\mathbf{r})}(\rho(\mathbf{r}')-\rho_{i\sigma}(\mathbf{r}') ) 
+ \frac{\rho(\mathbf{r}) -\rho_{i\sigma}(\mathbf{r})}{\rho(\mathbf{r})} 
h^{\text{LSD}}_{\text{xc}}(\mathbf{r},\mathbf{r}'; [\rho-\rho_{i\sigma}]) . 
\end{equation}
Turning to the second term and including the appropriate sign, one obtains
\begin{eqnarray}
h_{\text{xc},i\sigma}^{\rm (b)}(\mathbf{r},\mathbf{r}') &=& -\rho(\mathbf{r}') 
-h^{\text{LSD}}_{\text{xc}}(\mathbf{r},\mathbf{r}'; [\rho]) .
\end{eqnarray}
The third term is more complicated to derive since its expression is based on the exchange-correlation potential.
Making use of the relation
\begin{eqnarray}
v_{\text{xc},\sigma}(\mathbf{r}; [\rho]) = \frac 12 \int d\mathbf{r}'  \frac{ h_{\text{xc}}(\mathbf{r},\mathbf{r}'; [\rho])}{|\mathbf{r}-\mathbf{r}'|}
+ \frac{1}{2} \int d\mathbf{r}' d\mathbf{r}'' \frac{\rho(\mathbf{r}')}{|\mathbf{r}'-\mathbf{r}''|}
\frac{\delta h_{\text{xc}}(\mathbf{r}',\mathbf{r}''; [\rho])}{\delta \rho_{\sigma}(\mathbf{r})},
\end{eqnarray}
the expression for $h_{\text{xc},i\sigma}^{\rm (c)}$ becomes
\begin{equation}
h_{\text{xc},i\sigma}^{\rm (c)}(\mathbf{r},\mathbf{r}') =  
\frac{2\rho_{i\sigma}(\mathbf{r})}{\rho(\mathbf{r})}
\rho_{i\sigma}^{\text{ref}} (\mathbf{r}')
+\frac{\rho_{i\sigma}(\mathbf{r})}{\rho(\mathbf{r})}
h^{\text{LSD}}_{\text{xc}}(\mathbf{r},\mathbf{r}'; [\rho_{i\sigma}^{\text{ref}}])
+\frac{\rho^{\text{ref}}_{i\sigma}(\mathbf{r})}{\rho(\mathbf{r})}
\int d\mathbf{r}'' \rho_{i\sigma}(\mathbf{r}'')\frac{ \delta h^{\text{LSD}}_{\text{xc}}(\mathbf{r},\mathbf{r}';[\rho_{i\sigma}^{\text{ref}}]) }
{ \delta \rho_{\sigma}(\mathbf{r}'')}. 
\end{equation}

Regrouping all the terms, the expression for the exchange-correlation
hole of the $\alpha$NK functional can be written as
\begin{multline}
h^{\alpha \text{NK}}_{\text{xc}}(\mathbf{r},\mathbf{r}') = 
h^{\text{LSD}}_{\text{xc}}(\mathbf{r},\mathbf{r}'; [\rho])  
+\frac{\alpha^{\text{NK}}}{\rho(\mathbf{r})}\sum_{i \sigma} \Big(
\rho_{i\sigma}(\mathbf{r})\rho(\mathbf{r}') -\rho(\mathbf{r})\rho_{i\sigma}(\mathbf{r}')
+ \rho^{\text{ref}}_{i\sigma}(\mathbf{r})\int d\mathbf{r}'' \rho_{i\sigma}(\mathbf{r}'') 
\frac{ \delta h^{\text{LSD}}_{\text{xc}}(\mathbf{r},\mathbf{r}'; [\rho_{i\sigma}^{\text{ref}}]) }
{ \delta \rho_{\sigma}(\mathbf{r}'')} \\
+f_{i\sigma}(2f_{\text{ref}}-f_{i\sigma}) n_{i\sigma}(\mathbf{r})n_{i\sigma}(\mathbf{r}') 
+ h^{\text{LSD}}_{\text{xc}}(\mathbf{r},\mathbf{r}'; [\rho-\rho_{i\sigma}])(\rho(\mathbf{r})-\rho_{i\sigma}(\mathbf{r}))
- h^{\text{LSD}}_{\text{xc}}(\mathbf{r},\mathbf{r}'; [\rho])\rho(\mathbf{r}) 
+ h^{\text{LSD}}_{\text{xc}}(\mathbf{r},\mathbf{r}'; [\rho^{\text{ref}}_{i\sigma}])\rho_{i\sigma}(\mathbf{r})\Big). 
\end{multline}
The result given in Eq.~(\ref{xcHoleSumRuleNK}) can be obtained from the above equation
taking into account the xc-hole sum rule of the LSD functional, 
$\int d\mathbf{r}' h^{\text{LSD}}_{\text{xc}}(\mathbf{r},\mathbf{r}') = -1$ 
(valid for any electron number).

\end{widetext}

\bibliography{article}

\end{document}